\documentclass[prl,twocolumn,aps,showpacs,superscriptaddress]{revtex4}

\usepackage{graphicx}
\usepackage{dcolumn}
\usepackage{bm}
\usepackage{amsmath}

\begin{document}
\title{The Josephson light-emitting diode}

\author{Patrik Recher}
\altaffiliation{Current address: 
Institute for Theoretical Physics and Astrophysics, 
University of W\"urzburg, Germany }
\affiliation{Instituut-Lorentz, Universiteit Leiden, P.O. Box 9506, 2300 RA Leiden, The Netherlands}
\affiliation{Kavli Institute of Nanoscience, Delft University of Technology, P.O. Box 5046, 
2600 GA Delft, The Netherlands}

\author{Yuli V. Nazarov}
\affiliation{Kavli Institute of Nanoscience, Delft University of Technology, P.O. Box 5046, 
2600 GA Delft, The Netherlands}

\author{Leo P. Kouwenhoven} 
\affiliation{Kavli Institute of Nanoscience, Delft University of Technology, P.O. Box 5046, 
2600 GA Delft, The Netherlands}

\begin{abstract}
We consider an optical quantum dot where an electron level and a hole level are coupled to respective superconducting leads.
We find that electrons and holes recombine producing photons at discrete energies as well as a continuous tail. Further, the spectral lines directly probe the induced superconducting correlations on the dot. At energies close to the applied bias voltage $eV_{\rm sd}$, a parameter range exists, where radiation proceeds in pairwise emission of polarization correlated photons. At energies close to $2eV_{\rm sd}$, emitted photons are associated with Cooper pair transfer and are reminiscent of Josephson radiation. We discuss how to probe the coherence of these photons in a SQUID geometry via single photon interference.
\end{abstract}

\date{February 25, 2009}
\pacs{73.40.-c, 74.45.+c, 74.50.+r, 78.67.-n}
\narrowtext\maketitle
Electron-hole recombination in semiconductors accompanied by emission
of visible light is a key element of many technologies. Semiconducting QDs have been proposed
to enhance these technologies by engineering the frequencies of radiation~\cite{QDold}. In the context of modern research, 
they have been considered as a controllable 
source of single~\cite{Michler,Santori01,Zwiller01} and entangled two-photon pairs~\cite{Benson,Burkard}. 
QDs allow for integration of photon-based technologies 
and solid state systems where electronic degrees of freedom are used
to represent quantum information (e.g. electron spins in quantum dots (QDs)~\cite{Loss98}, charge-~\cite{Makhlin99} and flux qubits~\cite{Mooij99} in superconducting (SC) circuits), combining the advantages of both. For quantum information purposes, it is crucial that indistinguishable optical photons or pairs of photons can be created on demand. The semiconducting QDs provide means to achieve this~\cite{Santori02}.

SC Josephson junctions can also be a source of coherent radiation. When the junction is biased with a voltage $V_{\rm sd}$, photons with frequency $\omega=2eV_{\rm sd}/\hbar$  are emitted corresponding to Cooper pair transfers between the two SC leads. This radiation is coherent since the Cooper pair transfers
are coherent owing to macroscopic phase coherence of SC condensates 
involved~\cite{Josephson}. 
The frequency of Josephson radiation is limited by the SC energy gap $\Delta\sim$ 1 meV, $\hbar\omega =2eV_{\rm sd}<4\Delta$. This is three orders of magnitude
away from the optical frequency range.

Many theoretical predictions (e.g.~\cite{induced})
promote the combination of SCs and semiconductors 
within a single nanostructure. This difficult technological problem
attracted attention for a long time~\cite{vanweesreview}.
Recent progress has been achieved with semiconductor nanowires.
SC field-effect transistor~\cite{Leo1} and Josephson effect~\cite{Leo2} 
in a semiconducting
QD have been experimentally confirmed.

\begin{figure}[h]
\begin{center}
\includegraphics[width=0.75\columnwidth]{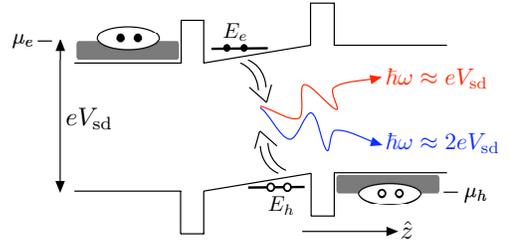}
\end{center}
\vspace{-0.4cm}
\caption{(Color online)
Sketch of a QD in contact to superconducting leads with chemical potentials $\mu_{e}$ and $\mu_{h}$. We consider a level in the conduction  band (with energy $E_{e}$)  and a level in the valence band (with energy $E_{h}$). 
Each level is only coupled to one of the reservoirs as indicated. Photon emission processes with energies $\hbar \omega$ close to the applied voltage bias $eV_{\rm sd}$ and at $2eV_{\rm sd}$ are indicated.}
\label{setup}
\vspace{-0.4cm}
\end{figure}

In this Letter, we propose and investigate theoretically a setup 
where a {\it superconducting p-n junction} enclosing 
a semiconducting QD emits photons in the optical range, see Fig.~1. 
This device is biased by a voltage $V_{\rm sd}$
which is close to the semiconducting band gap.
We show that, owing to SC correlations, 
the device emits the photons in the frequency range 
$eV_{\rm sd}/\hbar$ concentrated in several discrete spectral lines, 
the line-width being restricted by the emission time only.
The acts of photon emission correlate. In this way,
one can arrange emmision of pairs of photons of opposite polarization.
The device is also shown to emit in the frequency range $2eV_{\rm sd}/\hbar$.
The emitted light is associated with Cooper pair transfer between
the SC leads and is therefore coherent. This is in fact Josephson radiation at optical frequency. 

{\it Setup details}. 
The semiconducting QD encompasses two levels: 
one for electrons(e), one for holes(h). 
The levels are coupled to corresponding 
SC leads (source and drain), those being characterized
by energy gaps $\Delta_{e,h}$. The levels are 
aligned to the corresponding chemical potentials $\mu_{e,h}$.
We count their energies $E_{e,h}$ from these potentials
assuming $|E_{e,h}| \ll |\Delta_{e,h}|$.
The tunnel coupling in the normal state is characterized by
the broadening of a corresponding level,  
$\Gamma_{{\rm t};e,h}$, those being proportional to squares of the tunneling
amplitudes. In the presence of superconductivity,
we treat the coupling to the SC leads in second order 
perturbation theory~\cite{Supplementary}. This accounts for 
coherent transfers of electron singlets 
between the QD and the SC leads,
and amounts to an induced pair potential for the level, 
with ${\tilde \Delta}_{e,h}=(1/2)\exp[i\phi_{e,h}]\Gamma_{{\rm t};e,h}$
(assuming $\Gamma_{{\rm t};e,h} \ll |\Delta_{e,h}|$,~\cite{induced}),
and $\phi_{e,h}$ the phase of the corresponding
$\Delta_{e,h}$. 

The induced pair potential results in formation of
four discrete low-energy states at each 
(electron or hole) side of
the setup. We write the effective low-energy Hamiltonian
for electron side, skipping index ``e" for $\tilde\Delta$, $\Gamma$, $E$,
\begin{equation}
\label{g2}
{{\widetilde H}_{D}^{e}}=E\sum_{\sigma}c_{\sigma}^{\dagger}c_{\sigma}
+{\tilde \Delta} c_{\uparrow}^{\dagger}c_{\downarrow}^{\dagger}
+{\tilde \Delta}^{*}c_{\downarrow}c_{\uparrow}
+U {\hat n}_{\uparrow}{\hat n}_{\downarrow},
\end{equation}
where we assume that the charging energy 
(repulsive on-site interaction) $U \ll |\Delta|$.

By diagonalizing ${\widetilde H}_{D}^{e}$, we obtain two degenerate
single-particle states $
|\uparrow\rangle=c_{\uparrow}^{\dagger}|0\rangle$
and $
|\downarrow\rangle=c_{\downarrow}^{\dagger}|0\rangle$
with energy $E$ forming a doublet 
($|0\rangle$ denotes the empty level), 
and two singlets, those being linear superpositions
of $|0\rangle$ and $|2\rangle=c_{\uparrow}^{\dagger}c_{\downarrow}^{\dagger}|0\rangle$.
For the ground state singlet, we obtain
\begin{equation}
\label{g8m}
|g\rangle=-e^{-i\phi}|u|\,|0\rangle+|v|\,|2\rangle,
\end{equation}
with energy $\varepsilon_{g}={\tilde E}-({\tilde E}^2+|{\tilde \Delta}|^2)^{1/2} $, (${\tilde E}=E+(U/2)$). The coherence factors~\cite{Schrieffer} are
$|u|,|v|=(1/\sqrt{2})[1\pm {\tilde E}/({\tilde E}^{2}+|{\tilde \Delta}|^2)^{1/2}]^{1/2}$.
The excited state singlet reads
\begin{equation}
\label{g10m}
|e\rangle=e^{-i\phi}|v|\,|0\rangle+|u|\,|2\rangle,
\end{equation}
with energy $\varepsilon_{ex}={\tilde E}+
({\tilde E}^2+|{\tilde \Delta}|^2)^{1/2}.$
Similary, four states are formed on the hole side of
the setup. Since we are dealing with holes, we define
the corresponding vacuum $|0\rangle_{h}$ as the level
occupied by two electrons \cite{Supplementary}. Apart from this difference,
the energies and wave functions of the states are given
by above expressions with $E,\tilde\Delta,\Gamma_{\rm t},U
=E_h,\tilde\Delta_h,\Gamma_{{\rm t};h},U_h$. One could easily include
the interaction energy between electrons and holes in the above scheme.
We neglect this interaction since we do not expect it to change
our results qualitatively.

A SC p-n junction has been discussed in \cite{Hanamura}, and supplemented with a QD in \cite{Suemune06}, in the context of superradiance which is irrelevant for our proposed effects. 
\begin{figure}[h]
\begin{center}
\includegraphics[width=0.71\columnwidth]{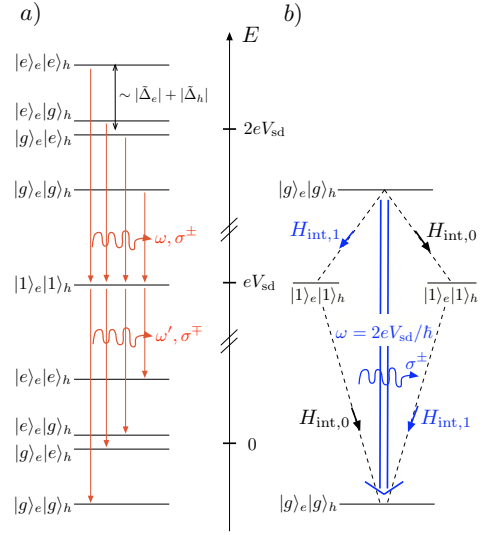}
\end{center}
\vspace{-0.4cm}
\caption{
(Color online) a) Recombination diagram of the biexciton-exciton cascade (EP cycle) with the QD coherently coupled to SC leads inducing 4 different singlet states and two (degenerate) doublet states for the combined system of electrons ($e$) and holes ($h$). The cascade produces two ``red'' photons with frequencies $\omega$ and $\omega'$ (of opposite circular polarization $\sigma^{\pm}$) on the order $eV_{\rm sd}/\hbar$. The cascade can proceed via 32 different decay paths (illustrated by red arrows) leading up to 8 distinct emission peaks as a consequence of induced gap ${\tilde \Delta}_{e,h}$ in the QD.
b) Illustration of ``blue" photon emission: The biexciton-exciton cascade can also proceed by emission of a {\it single} coherent photon
at the Josephson frequency $2eV_{\rm sd}$ which connects the same initial and final states (e.g. $|g\rangle_{e}|g\rangle_{h}$) differing by the transfer of one Cooper pair. This ``blue" photon can be ``stimulated" by an in-plane dc-electric field with Hamiltonian $H_{\rm int,0}$. The cascade then involves a ``zero-frequency photon" (non-radiative decay) as well as a ``blue" photon and the intermediate state $|1\rangle_{e}|1\rangle_{h}$ is only occupied virtually, whereas the total process conserves energy. }
\label{cycles}
\vspace{-0.3cm}
\end{figure}

{\it Emission of ``red" light}.
So far, we have not enabled charge transfer through the setup.
This can only proceed by recombination of an electron and a hole 
at different sides of the setup, see Fig.~\ref{setup}.
Such transfer has to dispose an energy $\simeq eV_{\rm sd}$ corresponding
the energy difference between  the electron and
hole level, and therefore is accompanied
by emission of a photon of this energy: Let us call it ``red" photon. 
The recombination is described by the following Hamiltonian:
\begin{equation}
\label{g13m}
H_{\rm int,1}=G\sum\limits_{q}\left(a_{q,-}^{\dagger}h_{\downarrow}c_{\uparrow}+a_{q,+}^{\dagger}h_{\uparrow}c_{\downarrow}\right)e^{-ieV_{{\rm sd}}t}+\rm{H.c.}
\end{equation}
The time dependence $\exp(\pm ieV_{{\rm sd}}t)$ accounts for the difference between $\mu_{e}$ and $\mu_{h}$.
Eq.~(\ref{g13m}) is a minimal model for the photon-assisted recombination 
of e-h pairs. We assume usual selection rules~\cite{Rochat} implying
that the holes are ``heavy", $h^{\dagger}_{\uparrow}$
($h^{\dagger}_{\downarrow}$) creates a hole with 
the total angular momentum $j_z = 3/2 (-3/2)$.
Eq.~(\ref{g13m}) then ensures the conservation of total angular momentum:
The polarization ($p=\pm$) of the photon emitted into a mode $q$
($a^{\dagger}_{q,\pm}$) is determined by the electron
and hole spins~\cite{Which_direction}.
An isolated QD in the state $|\uparrow\rangle_e|\downarrow\rangle_h$
would recombine to $|0\rangle_e|0\rangle_h$ with the rate 
$\Gamma_{\rm ph} \propto G^2$. Since the states of the
QD are modified by coupling to SC
leads [see Eqs.~(\ref{g8m}),(\ref{g10m})], the ``red" emission causes transitions 
between all QD states (see also Fig.~\ref{cycles}a)). 

An {\it even parity} emission cycle (EP) ($\# e+\#h$=even) and an {\it odd parity} (OP) cycle ($\# e+\#h$=odd) exist. 
The transitions  proceed
between the discrete states of the QD (see  Fig.~\ref{cycles}a)). 
They give rise
to sharp emission lines with
frequencies directly related to energy differences between the
states \cite{temperature}. The rates incorporate
the coherence factors, for instance,
\begin{eqnarray}
W^{p}_{|g\rangle_e|g\rangle_h \to |1\rangle_e |1\rangle_h} &=& (\Gamma_{\rm ph}/\hbar)
|v_e u_h|^2 ;\label{rate1} \\
W^{p}_{|1\rangle_e |1\rangle_h\to|g\rangle_e|g\rangle_h }  &=& (\Gamma_{\rm ph}/\hbar)
|u_e v_h|^2. \label{rate2}
\end{eqnarray}
EP and OP cycles are connected by transitions of a second type which involve the excitation of
a single quasiparticle with energy $>\Delta_{e,h}$ 
in one of the leads \cite{Supplementary} and therefore change 
the parity that is conserved in course of photon emission.
They give rise to a continuous spectrum of the ``red" light emitted
that is separated from the lines by frequency ${\rm min}(\Delta_e,\Delta_h)/\hbar$. The transition rates of the second type are smaller
as those of the first type by a typical reduction
factor $|\tilde\Delta|/|\Delta| \ll 1$. 

\begin{figure}[h]
\begin{center}
\includegraphics[width=0.82\columnwidth]{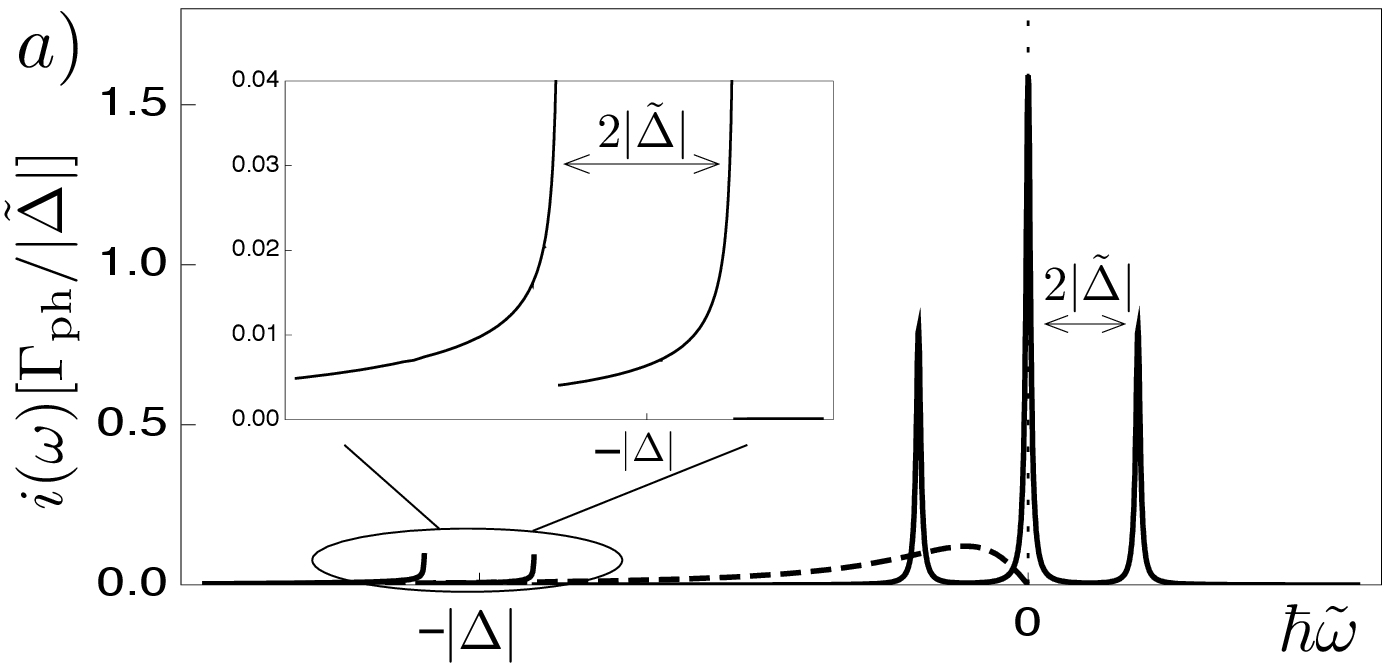}
\end{center}
\begin{center}
\includegraphics[width=0.66\columnwidth]{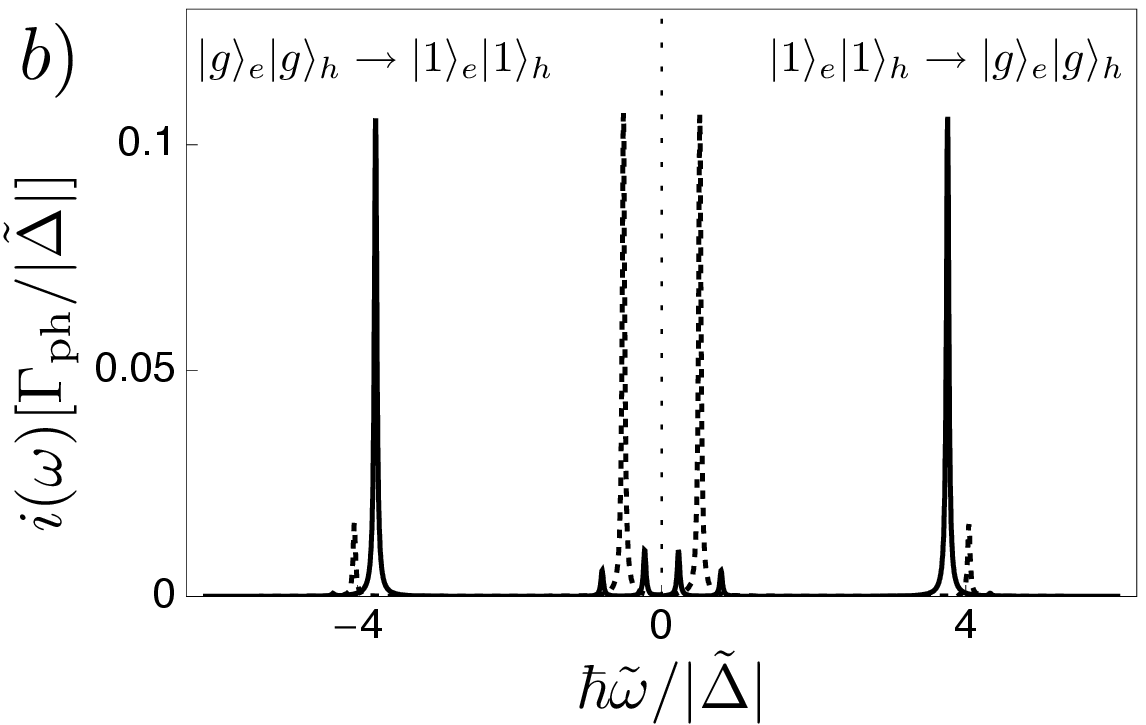}
\end{center}
\vspace{-0.5cm}
\caption{Emission intensity of ``red" photons in the energy range $|\hbar {\tilde\omega}|\equiv |\hbar \omega -eV_{\rm sd}|\sim |\Delta|$:
We use $|{\tilde \Delta}|$=0.1$|\Delta|$ and discrete peaks are broadened with $\Gamma_{\rm ph}$,
($\Gamma_{\rm ph}/\Gamma_{{\rm t}}=0.05$ a) and $0.02$ b)).
a) Spectrum at resonance $E_{e}=E_{h}=0$, $U=0$ encompasses 
several discrete peaks and a small continuous tail (inset). 
The dashed line shows intensity in the case of normal leads.  
b) Regime of pair-emission: Full lines and dotted lines show the emission spectrum from the EP- and OP cycle, resp.  Main full lines originate from time- and polarization correlated photons emitted from the biexciton-exciton cascade with groundstate singlets for electrons and holes. Parameters: $E_{e}=1.9$, $E_{h}=-1.6$, $U=0.28$ (in units of $|{\tilde \Delta}|$).}
\label{emission}
\vspace{-0.3cm}
\end{figure}

The emission intensity $i(\omega)=\sum_{ab,p}w^{p}_{a\rightarrow b}(\omega)\rho_{a}$ of the QD can be computed from the probabilities $\rho_a$ to be in one of 16 possible QD states $|a\rangle$. 
They follow from the stationary solution of the 
master equation describing the setup dynamics, governed by the rates
$W_{a\rightarrow b}^{p}=\int d\omega' \,w_{a\rightarrow b}^{p} (\omega')$~\cite{Supplementary}. 
The emission intensity computed is shown in Fig.~\ref{emission} versus photon frequency $\omega$ (we assume for simplicity that $|{\tilde \Delta}_{e}|=|{\tilde \Delta}_{h}|$ and $U_{e}=U_{h}$).
Plot a) gives the intensity
at the scale $|\hbar \omega- eV_{\rm sd}| \sim |\Delta|$
(for the case $E_{e}=E_{h}=U=0$). Three discrete peaks 
are visible at much smaller scale of the induced gap 
$|\hbar \omega- eV_{\rm sd}| \sim |\tilde\Delta|$. 
At  $\hbar\omega\approx eV_{\rm sd}-|\Delta|$, 
a continuous tail of emission starts (enlarged in the inset) reflecting quasiparticle creation in the leads. The dashed line is the emission spectrum of the same QD without superconductivity. In this case, the spectrum is
continuously broadened on the scale $\Gamma_{\rm t} = 2|\tilde\Delta|$. The total
emission intensity approximately corresponds to the total 
intensity of the three discrete lines in the SC case.  
Plot b) illustrates
the regime of {\it photon-pair} emission. 
The chosen parameters $E_{e}=1.9$, $E_{h}=-1.6 $ and $U=0.28 $ (in units of $|{\tilde \Delta}|$) induce a large population of the ground state singlet $|g\rangle_{e}|\rangle g\rangle_{h}$ ($\rho_{gg}\simeq 0.75$, $|u_{e}|\sim 0.97$ and $|v_{h}| \sim 0.96$). 
This has striking consequences for the cascade emission process  $|g\rangle_{e}|g\rangle_{h}\rightarrow |1\rangle_{e}|1\rangle_{h}\rightarrow |g\rangle_{e}|g\rangle_{h}$ shown in b) (main full lines). From Eqs.~(\ref{rate1}) and (\ref{rate2}) we deduce,  that $W_{|g\rangle_{e}|g\rangle_{h}\rightarrow|1\rangle_{e}|1\rangle_{h}}^{p}/W_{|1\rangle_{e}|1\rangle_{h}\rightarrow |g\rangle_{e}|g\rangle_{h}}^{p'}=|v_{e}u_{h}/u_{e}v_{h}|^2\ll 1$. Therefore, this process produces {\it two-photons} of {\it opposite polarization} in a pair (i.e. the delay time between the emission of the first and second photon is much shorter than the emission time of the pair) and with energies $\hbar\omega=eV_{\rm sd}\pm (\varepsilon_{g}^{e}+\varepsilon_{g}^{h}-E_{e}-E_{h})$. We point out that the energies of these correlated photons are different, however, the polarization and energy of the photons are {\it uncorrelated}. This cascade corresponds to the biexciton-exciton decay discussed in \cite{Benson} in the context of polarization-entangled photons. Therefore, potentially pairwise entangled photons \cite{entangledphotonsexp} could be identified efficiently in the time domain.
The dotted lines (OP cycle) are energetically distinct from the full lines (EP cycle) as a consequence of induced $|{\tilde \Delta}|$ and $U$. This allows us to distinguish emission processes from different cycles. We remark that the charge current through the device just equals the emission intensity of ``red" photons. 

{\it Emission of ``blue" light at $2eV_{\rm sd}$}.
We now consider the emission of a {\it single-photon} per Cooper pair transfer through the QD. Since the Cooper pair charge is $2e$, the energy associated with its transfer is $2eV_{\rm sd}$. If a single photon is emitted by this process, it must have a frequency $\sim 2eV_{\rm sd}/\hbar$ which we call a ``blue" photon.
Since only one electron-hole pair recombines radiatively in the emission process, we need a {\it static} in-plane electric field ${\bm E}_{0}$ that annihilates the other pair (see Fig.~\ref{cycles}b)). This annihilation without emission is described by 
\begin{equation}
\label{g13b}
H_{\rm int,0}=\left(V_{0}^{+}h_{\downarrow}c_{\uparrow}+V_{0}^{-}h_{\uparrow}c_{\downarrow}\right)e^{-ieV_{{\rm sd}}t}+\rm{H.c.},
\end{equation}
with $V_{0}^{\pm}\propto E_{0,x}\mp iE_{0,y}$.
To second order in the total interaction Hamiltonian $H_{\rm int}=H_{\rm int,1}+H_{\rm int,0}$, the rate to emit a single ``blue" photon (with polarization $p=\pm$) is $W_{a\rightarrow b}^{p}=(2\pi/\hbar)|{\cal A}_{a\rightarrow b}^{p}|^{2}\delta(\varepsilon_{b}-\varepsilon_{a}+\hbar\omega-2eV_{\rm sd})$ between initial state $|a\rangle$ (with energy $\varepsilon_{a}$) and final state $|b\rangle$ (with energy $\varepsilon_{b}$) of the QD. 
\begin{figure}[h]
\begin{center}
\vspace{-0.5cm}
\hbox{\resizebox{3.8cm}{!}{\includegraphics{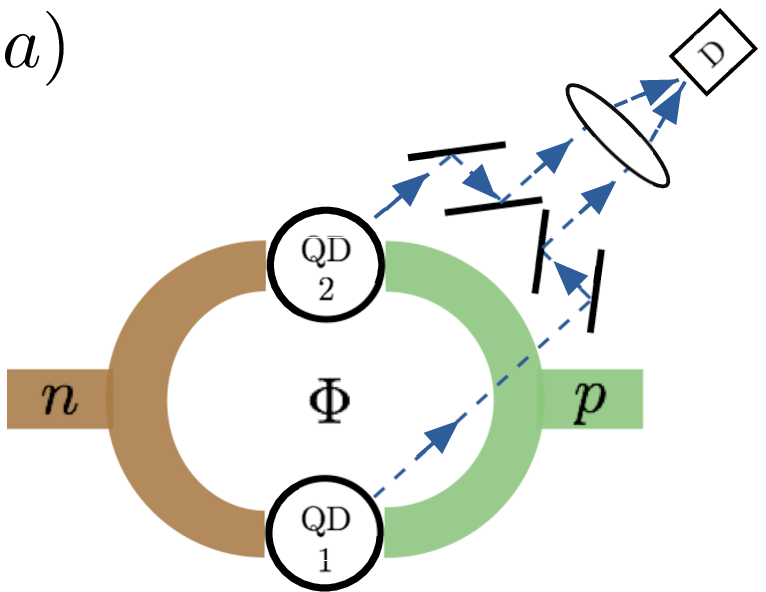}}\,\,\,\resizebox{4.6cm}{!}{\includegraphics{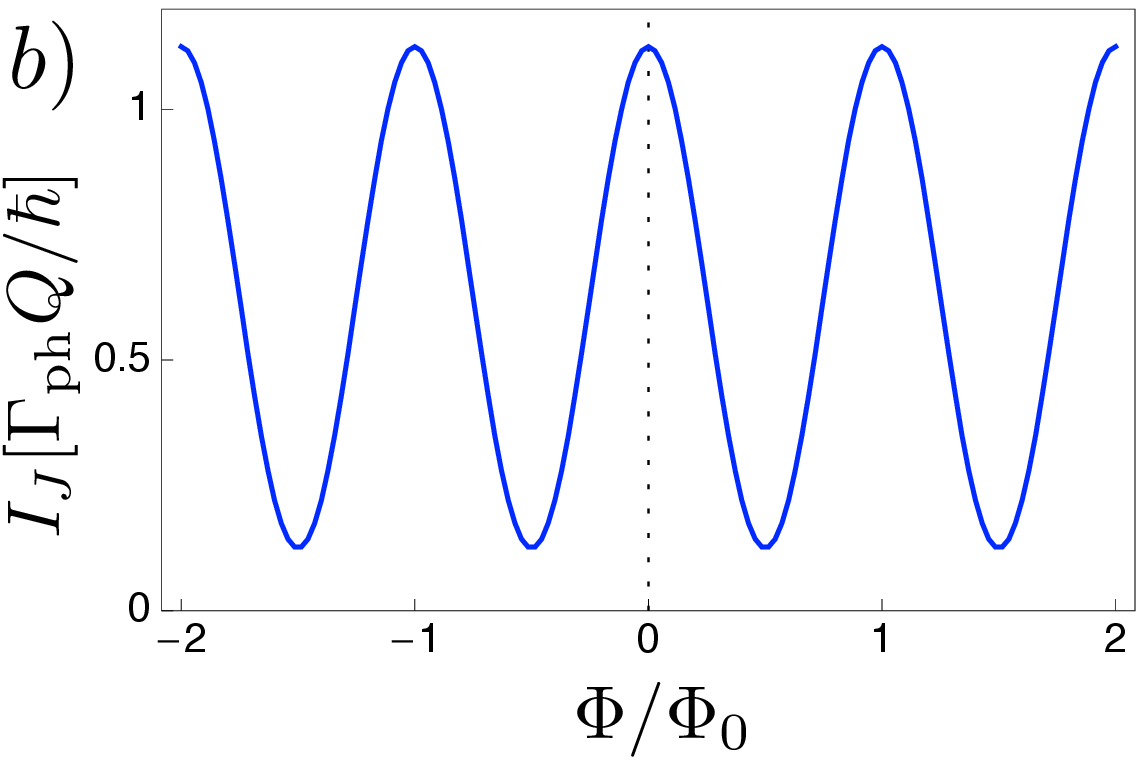}}}
\vspace{0.1cm}
\end{center}
\vspace{-0.5cm}
\caption{(Color online)
a) Proposed Josephson light-emitting diode in a SQUID configuration with two QDs (1,2) enclosing flux $\Phi$.
Interference of emitted photons at the Josephson frequency $2eV_{\rm sd}/\hbar$ is observed via an optical mirror system and photodetector D.
b) Intensity $I_{J}$ of photons at D as function of $\Phi$ (in units of $\Phi_{0}=hc/2e$) through SQUID for $E_{e}=-E_{h}=|{\tilde \Delta}|$, $U=0$ in both QDs. We choose equal optical path lengths from QDs 1,2 to D.}
\label{SQUID}
\vspace{-0.3cm}
\end{figure}
For the case where $|a\rangle$  and  $|b\rangle$ belong to the singlet subspace~\cite{oddcyclecohnote}, we obtain
(in leading order in $1/eV_{\rm sd}$), 
\begin{equation}
\label{ampjosephson}
{\cal A}_{a\rightarrow b}^{p}=GV_{0}^{p} \langle b|00\rangle\langle 22|a\rangle
\frac{2(E_{e}+E_{h})-\varepsilon_{a}-\varepsilon_{b}}{(eV_{\rm sd})^2}.
\end{equation}
We note that the amplitude can also connect different initial and final QD states resulting in {\it incoherent} photons. However, they are emitted at different frequencies. The light emitted at $\hbar\omega=2eV_{\rm sd}$ is always coherent. 
In this case, ${\cal A}_{a\rightarrow b}^{p}={\cal A}_{a\rightarrow a}^{p}\equiv {\cal A}_{a}^{p}$ and  $\langle a|00\rangle\langle 22|a\rangle=\pm \exp[i(\phi_{e}-\phi_{h})]|u_{e}u_{h}v_{e}v_{h}|$. 
The ``blue" emission out of the doublet states  $|\uparrow\rangle_{e}|\downarrow\rangle_{h}$ and $|\downarrow\rangle_{e}|\uparrow\rangle_{h}$ is anomalously small with ${\cal A}_{a\rightarrow b}^{p}\propto (eV_{\rm sd})^{-3}$ and is irrelevant.

Let us consider two QDs embedded in a SQUID loop as shown in Fig.~\ref{SQUID}a). Coherent emission from either QD (1 or 2) into a common photonic mode (and with the same polarization) has amplitudes ${\cal A}_{1,a}^{p}$ and ${\cal A}_{2,a'}^{p}$ (assuming the QDs are in states $|a\rangle$ and $|a'\rangle$, resp.). The total intensity $I_{J}$ of photons in the common mode is proportional to $\sum_{aa',p}\rho_{a}\rho_{a'}|{\cal A}_{1,a}^{p}e^{i2\pi l_1/\lambda_{J}}+ {\cal A}_{2,a'}^{p}e^{i2\pi l_2/\lambda_{J}}|^2$, where $\lambda_{J}=hc/2eV_{\rm sd}$ is the wave length of coherent light at the Josephson frequency and $l_{1}$ and $l_{2}$ are the respective path lengths from the QDs to the detector. The interference contribution is proportional to $\sum_{aa',p}\rho_{a}\rho_{a'}{\rm Re}[{\cal A}_{1,a}^{p}({\cal A}_{2,a'}^{p})^{*}]$ with ${\rm Re}[{\cal A}_{1,a}^{p}({\cal A}_{2,a'}^{p})^{*}]\propto \cos[2\pi((l_1-l_2)/\lambda_{J}+\Phi/\Phi_{0})]$, where we use that  $\phi_{1e}-\phi_{1h}-(\phi_{2e}-\phi_{2h})=2\pi\Phi/\Phi_{0}$, with $\Phi$ the flux through the SQUID and $\Phi_{0}=hc/2e$ the SC flux quantum. 

Fig.~\ref{SQUID}b) shows the computed emission intensity of $2eV_{\rm sd}$ photons as a function of flux $\Phi$. We find that the intensity oscillates with period given by the superconducting flux quantum $\Phi_{0}=hc/2e$, and has a magnitude of order $(\Gamma_{\rm ph}/\hbar)Q$ with $Q\equiv 2|V_{0}^{p}{\tilde \Delta}|^2/(eV_{\rm sd})^4=2|{\bm d}\cdot{\bm E}_{0}|^2|{\tilde \Delta}|^2/(eV_{\rm sd})^4$, where ${\bm d}$ is the optical dipole moment~\cite{Supplementary} of the QD~\cite{Matrixelement}. 
The electric field ${\bm E}_{0}$ could be created by gates. Typical critical field strengths before quenching the photoluminescence of optical QDs are on the order of several Volts/$\mu$m \cite{quenching}. Taking $|{\bm d}_{\ ||}|$ on the order of the QD diameter $\sim 20$ nm and estimating $|\tilde \Delta|\lesssim 1$ meV (bounded by $|\Delta|$), we arrive at an intensity $I_{J}\sim$ 4 photons/s assuming $eV_{\rm sd}\sim 1$eV and $\hbar/\Gamma_{\rm ph}\sim$ 0.1 ns at $2eV_{\rm sd}$. This intensity is measurable with single-photon detectors \cite{Lvovsky}. In addition, the Purcell effect in a QD-cavity system could enhance $\Gamma_{\rm ph}$ substantially  \cite{Purcell}. 

In conclusion, we investigated emission from a quantum dot (QD) embedded in a superconducting (SC) p-n junction. 
The presence of SC leads induces an effective pair-potential for electrons (e) and holes (h) on the QD. 
At frequencies $\omega$ close to the voltage bias $eV_{\rm sd}/\hbar$ of the p-n junction, a regime exists where radiation is correlated in pairs of oppositely polarized photons. 
At $\omega=2eV_{\rm sd}/\hbar$, emission is associated with Cooper pair transfer and is coherent. We proposed an experiment where interference of radiation from distant QDs arranged in a SQUID geometry can be manipulated by a magnetic flux. This provides a fascinating new tool to manipulate coherent light at optical frequencies. 

 We acknowledge useful discussions with N. Akopian, C.W.J. Beenakker, S. Frolov, D. Loss, U. Perinetti, and V. Zwiller and financial support from the Dutch Science Foundation NWO/FOM.

\newpage
\cleardoublepage
\setcounter{figure}{0}
\setcounter{equation}{0}
\section{Supplementary material for ``The Josephson light-emitting diode"}
\subsection{Spin-degenerate level coupled to a superconductor: effective Hamiltonian}
In this section we derive the effective Hamiltonian Eq.~(1) in the main text for a level (conduction or valence band) of the quantum dot (QD) coupled to a superconductor (SC).
The Hamiltonian for SC and the QD level coupled by tunneling is
$H=H_{S}+H_{D}+H_T$. The s-wave superconductor is described by the BCS Hamiltonian \cite{Tinkham}
\begin{equation}
H_{S}=\sum_{{\bf k}\sigma}\xi_{{\bf k}}c_{{\bf k}\sigma}^{\dagger}c_{{\bf k}\sigma}+\sum_{\bf k}\left(\Delta\ c_{{\bf k}\uparrow}^{\dagger}c_{-{\bf k}\downarrow}^{\dagger}+{\rm H.c.}\right),
\end{equation}
with $\xi_{k}=\varepsilon_{k}-\mu$ the single-particle energies in SC counted from the chemical potential $\mu$.
This Hamiltonian is diagonalized by the canonical transformation $c_{{\bf k}\uparrow}=u_{{k}}^{*}\gamma_{{\bf k}\uparrow}+v_{ k}\gamma_{-{\bf k}\downarrow}^{\dagger}$ and $c_{-{\bf k}\downarrow}=u_{k}^{*}\gamma_{-{\bf k}\downarrow}-v_{k}\gamma_{{\bf k}\uparrow}^{\dagger}$ and reads 
$H_S=\sum_{{\bf k}\sigma} E_{k}\gamma_{{\bf k}\sigma}^{\dagger}\gamma_{{\bf k}\sigma}$ with $E_{k}=\sqrt{\xi_{k}^2+|\Delta|^2}$, and $\Delta$ the superconducting pair-potential. The {\it isolated} QD is represented by  $H_{D}=E\sum_{\sigma}c_{\sigma}^{\dagger}c_{\sigma}+ U n_{\uparrow}n_{\downarrow}$, ($\sigma=\uparrow,\downarrow$) with $U$ a possible repulsive on-site interaction and $E$ the spin-degenerate energy level (counted from $\mu$). The tunneling Hamiltonian has the form $H_{T}=\sum_{{\bf k}\sigma}t_{\bf k}c_{\sigma}^{\dagger}c_{{\bf k}\sigma}+{\rm H.c.}$
The first step is to integrate out the SC by deriving an effective QD Hamiltonian in the subspace of the BCS groundstate taking into account the tunneling between the QD and SC. Defining $P$ as the projection operator for states of the total system with no excitations in SC, i.e. $\gamma_{k\sigma}P\psi=0$ for any state $\psi$, the effective Hamiltonian is \cite{Hubbard book}
\begin{equation}
\label{g1}
{\widetilde H_{D}}(\varepsilon)=PHP+PH\frac{1}{\varepsilon-QH}QHP,
\end{equation}
with $Q={\hat 1}-P$.
The first term on the RHS of Eq.~(\ref{g1}) is replaced by $H_{D}$ since the tunneling $H_{T}$ cannot act to first order in the subspace with projector $P$. To second order in $H_{T}$, the resonant transport of electron singlets between the QD and the SC is possible and described by the second term on the RHS of Eq.~(\ref{g1}). 
To leading order in $H_{T}$, this gives $PH(\varepsilon-QH)^{-1}QHP=PH_{T}[\varepsilon-(H_{S}+H_{D})]^{-1}H_{T}P$. The virtual energy cost $\varepsilon-Q(H_{D}+H_{S})$ (created by hopping of a single electron from (to) the QD to (from) the SC) with a quasiparticle of energy $E_{k}$ in the SC is approximated by $-E_{k}$ since we assume that $\Delta$ is the largest energy scale, i.e. $|\Delta| \gg |E|,U,t_{k}$. By tunneling of another electron (with opposite spin) from the QD to the SC (or vice versa), the excitation in SC can be removed and a Cooper pair is added (or removed) to (from) the condensate. These processes lead to the following contribution 
\begin{multline}
\label{g2}
PH_{T}[\varepsilon-(H_{S}+H_{D})]^{-1}H_{T}P\\
\sim -\sum_{{\bf k}}|t_{{\bf k}}|^2\frac{v_{{\bf k}}^{*}u_{{\bf k}}}{E_{k}}\left(c_{\downarrow}c_{\uparrow}-c_{\uparrow}c_{\downarrow}\right)+\rm{H.c.},
\end{multline}
where we used that $t_{-{\bf k}}^{*}=t_{{\bf k}}$. Since we are dealing with two SCs (electron-side and hole-side of the setup), it is important to keep track of the SC condensate phase $\phi$ which is related to the phases of the SC coherence factors \cite{Tinkham}: $v_{\bf k}^{*}u_{\bf k}=-|v_{k}|u_{k}|\exp(-i\phi)$, where $|u_{k}|=(1/\sqrt{2})(1+\xi_{k}/E_{k})^{1/2},$ $|v_{k}|=(1/\sqrt{2})(1-\xi_{k}/E_{k})^{1/2}$. By replacing the momentum sum in Eq.~(\ref{g2}) by an integral over energy, we obtain the following effective QD Hamiltonian
\begin{equation}
\label{g3}
{\widetilde H_{D}}=E\sum_{\sigma}c_{\sigma}^{\dagger}c_{\sigma}+{\tilde \Delta} c_{\uparrow}^{\dagger}c_{\downarrow}^{\dagger}+{\tilde \Delta}^{*}c_{\downarrow}c_{\uparrow}+U n_{\uparrow}n_{\downarrow},
\end{equation}
with ${\tilde \Delta}=(1/2)\exp(i\phi)\Gamma_{\rm t}$, where the level broadening $\Gamma_{\rm t}=2\pi\nu_{S}|t|^2$. Here, $\nu_{S}$ is the normal-state DOS per spin at the Fermi level $\mu$ in SC. Since the main part of the integral in Eq.~(\ref{g2}) comes from energies $|\xi_{k}|\lesssim |\Delta|\ll \varepsilon_{F}$ ($ \varepsilon_{F}$ the Fermi energy in the leads), the exact $k$-dependence of $t_{{\bf k}}$ can be neglected. 

\subsection{Energy levels of the QD coupled to SC leads}
The diagonalization of ${\tilde H_{D}}$ leads to four states for electrons and holes, see Fig.~1. For the electron side of the setup (with bare level energy $E_{e}$, induced gap ${\tilde \Delta}_{e}$ and on-site repulsion $U_{e}$) there is one doublet state
\begin{equation}
\label{g6}
|\uparrow\rangle_{e}=c_{\uparrow}^{\dagger}|0\rangle_{e}
\end{equation}
and 
\begin{equation}
\label{g7}
|\downarrow\rangle_{e}=c_{\downarrow}^{\dagger}|0\rangle_{e},
\end{equation}
with energy $E_{e}$, and two singlets (being a superpositions of zero and two electrons)
\begin{equation}
\label{g8}
|g\rangle_{e}=-e^{-i\phi_{e}}|u_{e}|\,|0\rangle_{e}+|v_{e}|\,|2\rangle_{e},
\end{equation}
with
\begin{equation}
\label{g9}
\varepsilon_{g}^{e}={\tilde E_{e}}-\sqrt{{\tilde E_{e}}^2+|{\tilde \Delta}_{e}|^2},
\end{equation}
where ${\tilde E_{e}}=E_{e}+U_{e}/2$, $|2\rangle_{e}=c_{\uparrow}^{\dagger}c_{\downarrow}^{\dagger}|0\rangle_{e}$ and $|0\rangle_{e}$ denotes the empty level. We have introduced the coherence factors $|u_{e}|=(1/\sqrt{2})[1+{\tilde E_{e}}/({\tilde E_{e}}^{2}+|{\tilde \Delta}_{e}|^2)^{1/2}]^{1/2}$ and $|v_{e}|=(1/\sqrt{2})[1-{\tilde E_{e}}/({\tilde E_{e}}^{2}+|{\tilde \Delta}_{e}|^2)^{1/2}]^{1/2}$.
The excited state involving the superconductor is
\begin{equation}
\label{g10}
|e\rangle_{e}=e^{-i\phi_{e}}|v_{e}|\,|0\rangle_{e}+|u_{e}|\,|2\rangle_{e},
\end{equation}
with
\begin{equation}
\label{g11}
\varepsilon_{ex}^{e}={\tilde E_{e}}+\sqrt{{\tilde E_{e}}^2+|{\tilde \Delta}_{e}|^2}.
\end{equation}
\begin{figure}[h]
\begin{center}
\includegraphics[width=0.7\columnwidth]{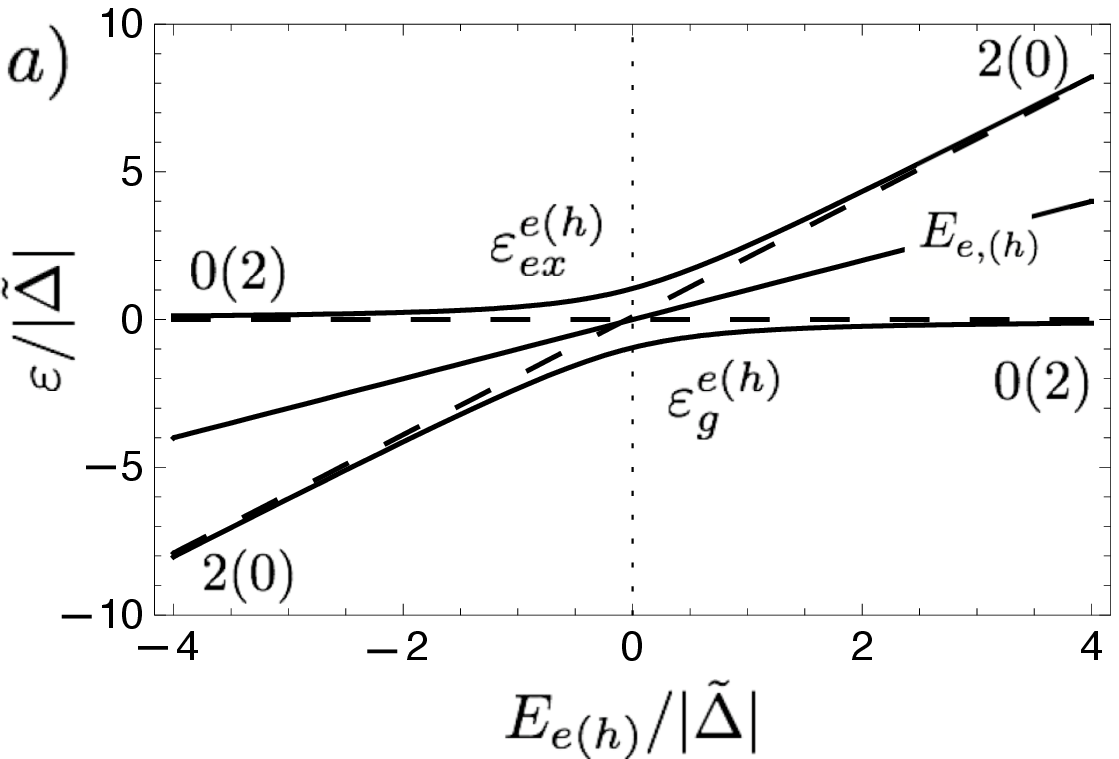}
\end{center}
\begin{center}
\includegraphics[width=0.7\columnwidth]{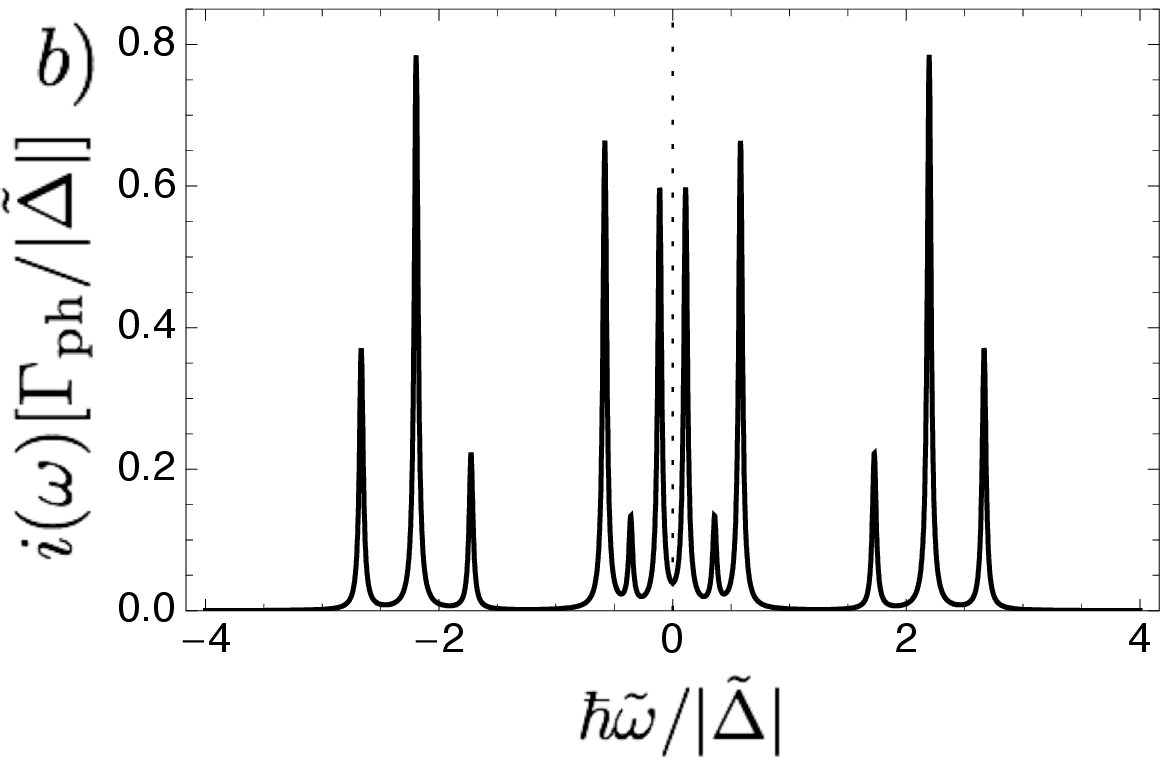}
\end{center}
\vspace{-0.4cm}
\caption{
a) Level structure of the QD coupled to SC leads: The coupling to SC leads induces four states and three distinct energies for electrons ($e$) and for holes ($h$). A doublet with energy $E_{e(h)}$ and two singlets with energies $\varepsilon_{g}^{e(h)}$ and $\varepsilon_{ex}^{e(h)}$ being superpositions of zero and two electrons(holes). The mixing character can be tuned by the QD level energy  $E_{e(h)}$. Away from resonance, $|E_{e(h)}|\gg |{\tilde \Delta}|$, the states become pure number states (shown with numbers for electrons(holes), energies given by dashed lines), where ${\tilde \Delta}$ is the induced SC gap in the QD. The level on-site energy is chosen as $U=0.1 |{\tilde \Delta}|$. b) Emission intensity of ``red" photons at $\hbar\omega\simeq eV_{\rm sd}$. In the generic case shown ($E_{e}=-0.81$, $E_{h}=-0.53$ and $U=0.47$ (in units of $|{\tilde \Delta}|$)) 12 peaks are visible (assuming $|{\tilde \Delta}_{e}|=|{\tilde \Delta}_{h}|\equiv |{\tilde \Delta}|$ and $U_{e}=U_{h}\equiv U$). Peaks are broadened with $\Gamma_{\rm ph}$ ($\Gamma_{\rm ph}/\Gamma_{t}=0.02$). }
\label{setup}
\end{figure}

For the hole-side of the setup the same four levels result (with $E_{e}\rightarrow E_{h}$, ${\tilde \Delta}_{e}\rightarrow {\tilde \Delta}_{h}$, and $U_{e}\rightarrow U_{h}$, $\phi_{e}\rightarrow \phi_{h}$). We then transform to the hole-picture for the valence band, by defining $|0\rangle_{h}=|2\rangle_{e}$ and $c_{\sigma}=h_{-\sigma}^{\dagger}$. Explicitly, the four levels on the hole-side are
\begin{equation}
\label{g12}
|\uparrow\rangle_{h}=h_{\uparrow}^{\dagger}|0\rangle_{h}
\end{equation}
and 
\begin{equation}
\label{g13}
|\downarrow\rangle_{h}=h_{\downarrow}^{\dagger}|0\rangle_{h},
\end{equation}
\begin{equation}
|g\rangle_{h}=-e^{-i\phi_{h}}|u_{h}|\,|2\rangle_{h}+|v_{h}|\,|0\rangle_{h}
\end{equation}
and
\begin{equation}
|e\rangle_{h}=e^{-i\phi_{h}}|v_{h}|\,|2\rangle_{h}+|u_{h}|\,|0\rangle_{h},
\end{equation}
with $|2\rangle_{h}=h_{\uparrow}^{\dagger}h_{\downarrow}^{\dagger}\,|0\rangle_{h}$.
\begin{figure}[h]
\vspace{0.6cm}
\begin{center}
\includegraphics[width=0.7\columnwidth]{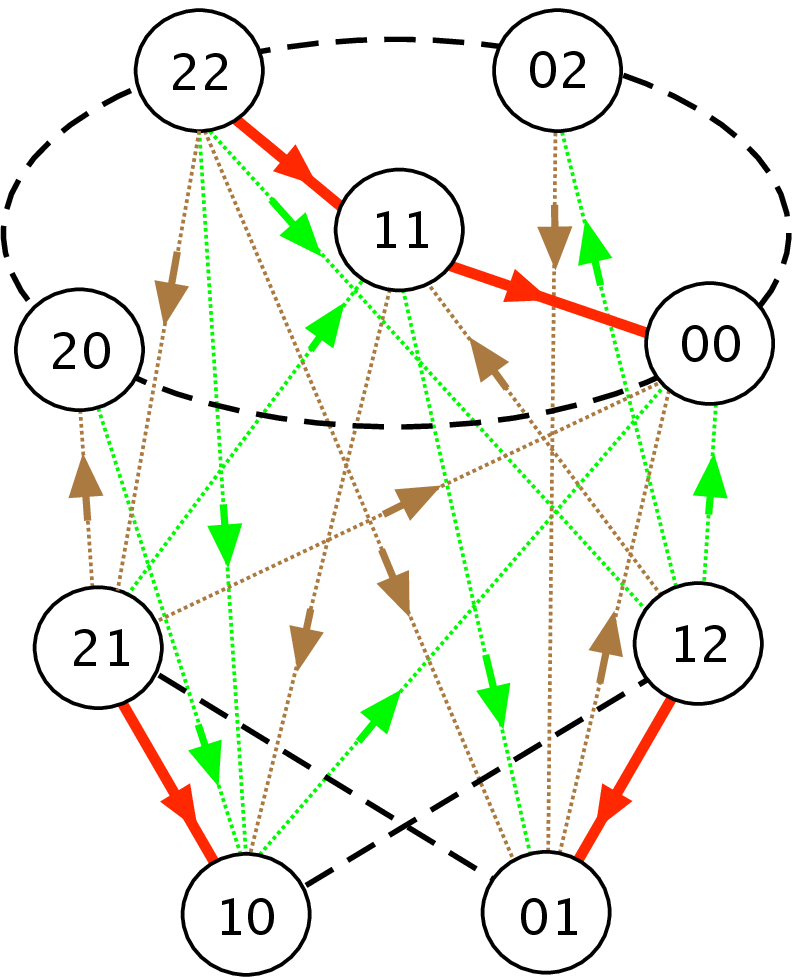}
\caption{(Color online) Setup dynamics and ``red" photon emission: QD number states and transitions between them involving emission of ``red" photons: Numbers within circles denote occupation numbers for electrons (left) and holes (right). Full lines (red) depict transitions (direction given by the arrow) where the total (e+h) number of particles on the levels change by two. Dashed black lines denote coherence between states mixed by SCs. Two such cycles exist: The upper one involves only states with an {\it even} number of total particles (even parity (EP) cycle) whereas the lower cycle has an {\it odd} number of total particles (odd parity (OP) cycle). The two cycles are connected by processes (dotted lines) which create a quasiparticle excitation in one of the leads (brown for $n$-side, green for $p$-side) in combination of emission of a ``red'' photon.}
\label{scheme1}
\end{center}
\end{figure}

\subsection{``Red" photon 
emission rates and master equation}
In this section we derive the rates for spontaneous emission of photons due to electron-hole recombination which 
is also responsible for charge transport through the QD. 
The interaction Hamiltonian with the elm-field takes on the form (Eq.~(4) of main text)
\begin{equation}
\label{g14}
H_{\rm int,1}=G\sum\limits_{q}\left(a_{q,-}^{\dagger}h_{\downarrow}c_{\uparrow}+a_{q,+}^{\dagger}h_{\uparrow}c_{\downarrow}\right)e^{-ieV_{{\rm sd}}t}+\rm{H.c.}
\end{equation}
The circular polarization ($p=\pm$) of the photon emitted into a mode $q$
($a^{\dagger}_{q,\pm}$) is determined by the electron
and hole spins \cite{ph_direction}, and the Hamiltonian for photons is $H_{\rm ph}=\sum_{q,p=\pm}\hbar\omega_{q} a_{q,p}^{\dagger}a_{q,p}$.

We treat the interaction Hamiltonian with the elm-field $H_{\rm int,1}$ as a perturbation and assume $\Gamma_{\rm ph}\ll |{\tilde\Delta_{e.h}}|$, where $\Gamma_{\rm ph}=2\pi\nu_{\rm ph}|G|^2$ with $\nu_{\rm ph}$ the photon DOS per polarization direction, assumed to be independent of energy and polarization. 
We use a stationary master equation approach to calculate the occupation probability $\rho_{a}$ for each of the 16 possible states $|a\rangle$ of the combined system of electrons and holes. These states $|a\rangle$ are: $|g\rangle_{e}|g\rangle_{h}$, $|e\rangle_{e}|e\rangle_{h}$, $|g\rangle_{e}|e\rangle_{h}$, $|e\rangle_{e}|g\rangle_{h}$, $|g\rangle_{e}|\uparrow\rangle_{h}$, $|g\rangle_{e}|\downarrow\rangle_{h}$, $|e\rangle_{e}|\uparrow\rangle_{h}$, $|e\rangle_{e}|\downarrow\rangle_{h}$, $|\uparrow\rangle_{e}|g\rangle_{h}$, $|\downarrow\rangle_{e}|g\rangle_{h}$, $|\uparrow\rangle_{e}|e\rangle_{h}$, $|\downarrow\rangle_{e}|e\rangle_{h}$, $|\uparrow\rangle_{e}|\downarrow\rangle_{h}$, $|\downarrow\rangle_{e}|\uparrow\rangle_{h}$, $|\uparrow\rangle_{e}|\uparrow\rangle_{h}$, $|\downarrow\rangle_{e}|\downarrow\rangle_{h}$. 

The states are connected by rates of the form $W_{b,a}^{p}$, where $|a\rangle$ is the initial state of the QD (energy $\varepsilon_{a}$) and $|b\rangle$ is the final state of the QD (energy $\varepsilon_{b}$) via the emission of a photon with energy $\hbar\omega$ and polarization $p=\pm$. They are given by the usual form \cite{Sakurai}
\begin{equation}
W_{b,a}^{p}=\frac{2 \pi}{\hbar} \sum_{q}\left|\langle b;q,p| H_{\rm int,1} |a;0\rangle\right|^{2}\,\delta(\varepsilon_{a}-\varepsilon_{b}-\hbar{\tilde \omega}_{q}),
\end{equation}
with ${\tilde \omega}\equiv \omega-eV_{\rm sd}/\hbar$. 
The dynamics of the system is illustrated by the diagram in Fig.~2. 
Two emission cycles exist:  A cycle where $\#e+\#h$ is even (upper cycle) which we refer to as the even parity (EP) cycle and a cycle where $\#e+\#h$ is odd (lower cycle) which we refer to as the odd parity (OP) cycle. Full red lines connect states within the same cycle (in the direction of arrows) via emission of a photon with energy $\simeq eV_{\rm sd}$. The two cycles are connected by rates of a second type (illustrated in Fig.~3 for a specific example). Since the parity cannot change in the course of photon emission only, these cycle connecting processes (depicted by dotted lines in Fig.~2) create an excitation in one of the leads via single-particle tunneling.
Since $eV_{\rm sd}\gg\Delta_{e,h}$ such processes are possible in combination with photon emission and will be discussed in more detail below.

Formally, the dynamics is governed by the following master equation 
(we abbreviate the QD states $|\alpha\rangle_{e}|\beta\rangle_{h}$ as $\alpha\beta$)

\begin{widetext}
\begin{multline}
\dot\rho_{gg}=W_{gg,\uparrow\downarrow}^{-}\,\rho_{\uparrow\downarrow}+W_{gg,\downarrow\uparrow}^{+}\,\rho_{\downarrow\uparrow}+W_{gg,\uparrow g}^{-}\,\rho_{\uparrow g}+W_{gg,\downarrow g}^{+}\,\rho_{\downarrow g}+W_{gg,g\uparrow}^{+}\,\rho_{g\uparrow}+W_{gg,g\downarrow}^{-}\,\rho_{g\downarrow}\\
-\left[W_{\uparrow\downarrow,gg}^{+}+W_{\downarrow\uparrow,gg}^{-}+W_{\uparrow g,gg}^{+}+W_{\downarrow g,gg}^{-}+W_{g\uparrow,gg}^{-}+W_{g\downarrow,gg}^{+}\right]\,\rho_{gg},
\end{multline}
\begin{multline}
\dot\rho_{ee}=W_{ee,\uparrow\downarrow}^{-}\,\rho_{\uparrow\downarrow}+W_{ee,\downarrow\uparrow}^{+}\,\rho_{\downarrow\uparrow}+W_{ee,\uparrow e}^{-}\,\rho_{\uparrow e}+W_{ee,\downarrow e}^{+}\,\rho_{\downarrow e}+W_{ee,e\uparrow}^{+}\,\rho_{e\uparrow}+W_{ee,e\downarrow}^{-}\,\rho_{e\downarrow}\\
-\left[W_{\uparrow\downarrow,ee}^{+}+W_{\downarrow\uparrow,ee}^{-}+W_{\uparrow e,ee}^{+}+W_{\downarrow e,ee}^{-}+W_{e\uparrow,ee}^{-}+W_{e\downarrow,ee}^{+}\right]\,\rho_{ee},
\end{multline}
\begin{multline}
\dot\rho_{ge}=W_{ge,\uparrow\downarrow}^{-}\,\rho_{\uparrow\downarrow}+W_{ge,\downarrow\uparrow}^{+}\,\rho_{\downarrow\uparrow}+W_{ge,\uparrow e}^{-}\,\rho_{\uparrow e}+W_{ge,\downarrow e}^{+}\,\rho_{\downarrow e}+W_{ge,g\uparrow}^{+}\,\rho_{g\uparrow}+W_{ge,g\downarrow}^{-}\,\rho_{g\downarrow}\\
-\left[W_{\uparrow\downarrow,ge}^{+}+W_{\downarrow\uparrow,ge}^{-}+W_{\uparrow e,ge}^{+}+W_{\downarrow e,ge}^{-}+W_{g\uparrow,ge}^{-}+W_{g\downarrow,ge}^{+}\right]\,\rho_{ge},
\end{multline}
\begin{multline}
\dot\rho_{eg}=W_{eg,\uparrow\downarrow}^{-}\,\rho_{\uparrow\downarrow}+W_{eg,\downarrow\uparrow}^{+}\,\rho_{\downarrow\uparrow}+W_{eg,\uparrow g}^{-}\,\rho_{\uparrow g}+W_{eg,\downarrow g}^{+}\,\rho_{\downarrow g}+W_{eg,e\uparrow}^{+}\,\rho_{e\uparrow}+W_{eg,e\downarrow}^{-}\,\rho_{e\downarrow}\\
-\left[W_{\uparrow\downarrow,eg}^{+}+W_{\downarrow\uparrow,eg}^{-}+W_{\uparrow g,eg}^{+}+W_{\downarrow g,eg}^{-}+W_{e\uparrow,eg}^{-}+W_{e\downarrow,eg}^{+}\right]\,\rho_{eg},
\end{multline}
\begin{multline}
\dot\rho_{g\uparrow}=W_{g\uparrow,\uparrow g}^{-}\,\rho_{\uparrow g}+W_{g\uparrow,\uparrow e}^{-}\,\rho_{\uparrow e}+W_{g\uparrow,gg}^{-}\,\rho_{gg}+W_{g\uparrow,ge}^{-}\,\rho_{ge}+W_{g\uparrow,\downarrow\uparrow}^{+}\,\rho_{\downarrow\uparrow}+W_{g\uparrow,\uparrow\uparrow}^{-}\,\rho_{\uparrow\uparrow}\\
-\left[W_{\uparrow g,g\uparrow}^{+}+W_{\uparrow e,g\uparrow}^{+}+W_{gg,g\uparrow}^{+}+W_{ge,g\uparrow}^{+}+W_{\downarrow\uparrow,g\uparrow}^{-}+W_{\uparrow\uparrow,g\uparrow}^{+}\right]\,\rho_{g\uparrow},
\end{multline}
\begin{multline}
\dot\rho_{g\downarrow}=W_{g\downarrow,\downarrow g}^{+}\,\rho_{\downarrow g}+W_{g\downarrow,\downarrow e}^{+}\,\rho_{\downarrow e}+W_{g\downarrow,gg}^{+}\,\rho_{gg} +W_{g\downarrow, ge}^{+}\,\rho_{ge}+W_{g \downarrow,\downarrow\downarrow}^{+}\,\rho_{\downarrow\downarrow} +W_{g\downarrow,\uparrow\downarrow}^{-}\,\rho_{\uparrow\downarrow}\\-\left[W_{\downarrow g,g\downarrow}^{-}+W_{\downarrow e,g\downarrow}^{-}+W_{ge,g\downarrow}^{-}+W_{gg,g\downarrow}^{-}+W_{\uparrow\downarrow,g\downarrow}^{+}+W_{\downarrow\downarrow,g\downarrow}^{-}\right]\,\rho_{g\downarrow},
\end{multline}
\begin{multline}
\dot\rho_{e\uparrow}=W_{e\uparrow,\uparrow g}^{-}\,\rho_{\uparrow g}+W_{e\uparrow,\uparrow e}^{-}\,\rho_{\uparrow  e}+W_{e\uparrow, ee}^{-}\,\rho_{ee}+W_{e\uparrow, eg}^{-}\,\rho_{eg}+W_{e\uparrow,\uparrow\uparrow}^{-}\,\rho_{\uparrow\uparrow}+W_{e\uparrow,\downarrow\uparrow}^{+}\,\rho_{\downarrow\uparrow}\\
-\left[W_{eg,e\uparrow}^{+}+W_{ee,e\uparrow}^{+}+W_{\uparrow g,e\uparrow}^{+}+W_{\uparrow e,e\uparrow}^{+}+W_{\uparrow\uparrow,e\uparrow}^{+}+W_{\downarrow\uparrow,e\uparrow}^{-}\right]\,\rho_{e\uparrow},
\end{multline}
\begin{multline}
\dot\rho_{e\downarrow}=W_{e\downarrow,\downarrow e}^{+}\,\rho_{\downarrow e}+W_{e\downarrow,\downarrow g}^{+}\,\rho_{\downarrow g}+W_{e\downarrow, ee}^{+}\,\rho_{ee}+W_{e\downarrow,eg}^{+}\rho_{eg}+W_{e\downarrow, \downarrow\downarrow}^{+}\,\rho_{\downarrow\downarrow}+W_{e\downarrow,\uparrow\downarrow}^{-}\,\rho_{\uparrow\downarrow}\\ -\left[W_{\downarrow g,e\downarrow}^{-}+W_{\downarrow e,e\downarrow}^{-}+W_{eg,e\downarrow}^{-}+W_{ee,e\downarrow}^{-}+W_{\downarrow\downarrow,e\downarrow}^{-}+W_{\uparrow\downarrow,e\downarrow}^{+}\right]\,\rho_{e\downarrow},
\end{multline}
\begin{multline}
\dot\rho_{\uparrow g}=W_{\uparrow g,e\uparrow}^{+}\,\rho_{e\uparrow}+W_{\uparrow g,g\uparrow}^{+}\,\rho_{g\uparrow}+W_{\uparrow g,gg}^{+}\,\rho_{gg}+W_{\uparrow g,eg}^{+}\,\rho_{eg}+W_{\uparrow g,\uparrow\uparrow}^{+}\,\rho_{\uparrow\uparrow}+W_{\uparrow g,\uparrow\downarrow}^{-}\,\rho_{\uparrow\downarrow}\\-\left[W_{g\uparrow,\uparrow g}^{-}+W_{e\uparrow,\uparrow g}^{-}+W_{eg,\uparrow g}^{-}+W_{gg,\uparrow g}^{-}+W_{\uparrow\downarrow,\uparrow g}^{+}+W_{\uparrow\uparrow,\uparrow g}^{-}\right]\,\rho_{\uparrow g},
\end{multline}
\begin{multline}
\dot\rho_{\downarrow g}=W_{\downarrow g,e\downarrow}^{-}\,\rho_{e\downarrow}+W_{\downarrow g,g\downarrow}^{-}\rho_{g\downarrow}+W_{\downarrow g,gg}^{-}\rho_{gg}+W_{\downarrow g,eg}^{-}\rho_{eg}+W_{\downarrow g,\downarrow\downarrow}^{-}\rho_{\downarrow\downarrow}+W_{\downarrow g,\downarrow\uparrow}^{+}\rho_{\downarrow\uparrow}\\-\left[W_{g\downarrow,\downarrow g}^{+}+W_{e\downarrow,\downarrow g}^{+}+W_{eg,\downarrow g}^{+}+W_{gg,\downarrow g}^{+}+W_{\downarrow\uparrow,\downarrow g}^{-}+W_{\downarrow\downarrow,\downarrow g}^{+}\right]\,\rho_{\downarrow g},
\end{multline}
\begin{multline}
\dot\rho_{\uparrow e}=W_{\uparrow e,g\uparrow}^{+}\,\rho_{g\uparrow}+W_{\uparrow e,e\uparrow}^{+}\,\rho_{e\uparrow}+W_{\uparrow e,ge}^{+}\,\rho_{ge}+W_{\uparrow e,ee}^{+}\,\rho_{ee}+W_{\uparrow e,\uparrow\uparrow}^{+}\,\rho_{\uparrow\uparrow}+W_{\uparrow e,\uparrow\downarrow}^{-}\,\rho_{\uparrow\downarrow}\\-\left[W_{g\uparrow,\uparrow e}^{-}+W_{e\uparrow,\uparrow e}^{-}+W_{ee,\uparrow e}^{-}+W_{ge,\uparrow e}^{-}+W_{\uparrow\downarrow,\uparrow e}^{+}+W_{\uparrow\uparrow,\uparrow e}^{-}\right]\,\rho_{\uparrow e},
\end{multline}
\begin{multline}
\dot\rho_{\downarrow e}=W_{\downarrow e,g\downarrow}^{-}\,\rho_{g\downarrow}+W_{\downarrow e,e\downarrow}^{-}\,\rho_{e\downarrow}+W_{\downarrow e,ge}^{-}\,\rho_{ge}+W_{\downarrow e,ee}^{-}\,\rho_{ee}+W_{\downarrow e,\downarrow\downarrow}^{-}\,\rho_{\downarrow\downarrow}+W_{\downarrow e,\downarrow\uparrow}^{+}\,\rho_{\downarrow\uparrow}\\-\left[W_{g\downarrow,\downarrow e}^{+}+W_{e\downarrow,\downarrow e}^{+}+W_{ee,\downarrow e}^{+}+W_{ge,\downarrow e}^{+}+W_{\downarrow\uparrow,\downarrow e}^{-}+W_{\downarrow\downarrow,\downarrow e}^{+}\right]\,\rho_{\downarrow e},
\end{multline}
\begin{multline}
\dot\rho_{\uparrow\downarrow}=W_{\uparrow\downarrow,gg}^{+}\,\rho_{gg}+W_{\uparrow\downarrow,ge}^{+}\,\rho_{ge}+W_{\uparrow\downarrow,eg}^{+}\,\rho_{eg}+W_{\uparrow\downarrow,ee}^{+}\,\rho_{ee}+W_{\uparrow\downarrow,\uparrow e}^{+}\,\rho_{\uparrow e}+W_{\uparrow\downarrow,g\downarrow}^{+}\,\rho_{g\downarrow}+W_{\uparrow\downarrow,e\downarrow}^{+}\,\rho_{e\downarrow}+W_{\uparrow\downarrow,\uparrow g}^{+}\,\rho_{\uparrow g}\\-\left[W_{gg,\uparrow\downarrow}^{-}+W_{ge,\uparrow\downarrow}^{-}+W_{eg,\uparrow\downarrow}^{-}\right. \left.+W_{ee,\uparrow\downarrow}^{-}+W_{\uparrow g,\uparrow\downarrow}^{-}+W_{\uparrow e,\uparrow\downarrow}^{-}+W_{g\downarrow,\uparrow\downarrow}^{-}+W_{e\downarrow,\uparrow\downarrow}^{-}\right]\,\rho_{\uparrow\downarrow},
\end{multline}
\begin{multline}
\dot\rho_{\downarrow\uparrow}=W_{\downarrow\uparrow,\downarrow e}^{-}\,\rho_{\downarrow e}+W_{\downarrow\uparrow,\downarrow g}^{-}\,\rho_{\downarrow g}+W_{\downarrow\uparrow,g\uparrow}^{-}\,\rho_{g\uparrow}+W_{\downarrow\uparrow,e\uparrow}^{-}\,\rho_{e\uparrow}+W_{\downarrow\uparrow,gg}^{-}\,\rho_{gg}+W_{\downarrow\uparrow,ge}^{-}\,\rho_{ge}
+W_{\downarrow\uparrow,eg}^{-}\,\rho_{eg}+W_{\downarrow\uparrow,ee}^{-}\,\rho_{ee}\\-\left[W_{\downarrow g,\downarrow\uparrow}^{+}+W_{\downarrow e,\downarrow\uparrow}^{+}+W_{g\uparrow,\downarrow\uparrow}^{+}+W_{e\uparrow,\downarrow\uparrow}^{+}\right.
\left.+W_{ge,\downarrow\uparrow}^{+}+W_{eg,\downarrow\uparrow}^{+}+W_{ee,\downarrow\uparrow}^{+}+W_{gg,\downarrow\uparrow}^{+}\right]\,\rho_{\downarrow\uparrow},
\end{multline}
\begin{multline}
\dot\rho_{\uparrow\uparrow}=W_{\uparrow\uparrow,\uparrow g}^{-}\,\rho_{\uparrow g}+W_{\uparrow\uparrow,\uparrow e}^{-}\,\rho_{\uparrow e}+W_{\uparrow\uparrow,g \uparrow}^{+}\,\rho_{g \uparrow}+W_{\uparrow\uparrow,e\uparrow}^{+}\,\rho_{e\uparrow} -\left[W_{\uparrow e,\uparrow\uparrow}^{+}+W_{\uparrow g,\uparrow\uparrow}^{+}+W_{g\uparrow,\uparrow\uparrow}^{-}+W_{e\uparrow,\uparrow\uparrow}^{-}\right]\,\rho_{\uparrow\uparrow},
\end{multline}
\begin{multline}
\dot\rho_{\downarrow\downarrow}=W_{\downarrow\downarrow,\downarrow g}^{+}\,\rho_{\downarrow g}+W_{\downarrow\downarrow,\downarrow e}^{+}\,\rho_{\downarrow e}+W_{\downarrow\downarrow,g\downarrow}^{-}\,\rho_{g\downarrow}+W_{\downarrow\downarrow,e\downarrow}^{-}\,\rho_{e\downarrow}-\left[W_{\downarrow e,\downarrow\downarrow}^{-}+W_{\downarrow g,\downarrow\downarrow}^{-}+W_{g\downarrow,\downarrow\downarrow}^{+}+W_{e\downarrow,\downarrow\downarrow}^{+}\right]\,
\rho_{\downarrow\downarrow}.
\end{multline}
\end{widetext}
The transition rates within the same cycle have the following form, e.g. 
\begin{equation}
\label{REP1}
W_{\downarrow\uparrow,gg}^{-}=(\Gamma_{\rm ph}/\hbar)|v_{e}u_{h}|^{2},
\end{equation}
which emits a $\sigma^{+}$-photon at energy $\hbar {\tilde \omega}=\varepsilon_{g}^{e}+\varepsilon_{g}^{h}-E_{e}-E_{h}$.\\
An example for the OP cycle is
\begin{equation}
\label{ROP1}
W_{g\downarrow,\downarrow e}^{+}=(\Gamma_{\rm ph}/\hbar)\,|u_{e}v_{h}|^{2},
\end{equation}
which emits a $\sigma^{+}$-photon at energy $\hbar{\tilde\omega}=\varepsilon_{ex}^{h}-\varepsilon_{g}^{e}+E_{e}-E_{h}$.

The rates that connect the two cycles involve the transition operator ${\hat V}(\varepsilon_{a}-{H}_{0})^{-1}{\hat V}$ with ${H}_{0}={{\widetilde H}}_{D}^{e}+{{\widetilde H}}_{D}^{h}+H_{S}+H_{\rm ph}$, ${\hat V}=H_{\rm int,1}+H_{T}^{e}+H_{T}^{h}$ and $\varepsilon_{a}$ is the energy of the QD before the transition (i.e. $|a\rangle$ is an eigenstate of $H_{0}$ with no quasiparticle in SC leads and no photons present). These rates are different since they involve the tunneling of an electron (hole) into/from the SC reservoirs (creating a {\it quasiparticle} with energy of at least $|\Delta_{e,h}|$), in combination with emission of a photon such that the total energy is conserved in the final state.
These processes have the following rates, e.g.
\begin{multline}
\label{cr1}
w_{\uparrow g,gg}^{+}(\omega,{\bm k}\downarrow)= \Gamma_{\rm ph}|t_{h}|^2|v_{e}|^2\\\times\frac{|u_{h}(\xi_{k})|^2}{E(\xi_{ k})^{2}}\,\delta\left(\hbar {\tilde \omega}-\varepsilon_{g}^{e}+E_{e}+E(\xi_{k})\right),
\end{multline}
which creates a $\sigma^{+}$-photon at energy $\hbar {\tilde \omega}=\varepsilon_{g}^{e}-E_{e}-E(\xi_{k})$ and a quasiparticle in the SC reservoir (hole-side) with spin down, momentum ${\bm k}$ and energy $E(\xi_{k})=\sqrt{\xi_{k}^2+|\Delta_{h}|^2}$, i.e. the state $\gamma_{h{\bm k}\downarrow}^{\dagger}|0\rangle_{\rm BCS}$, see Fig.~3. 
We can integrate over the quasiparticle state in the lead to get the rate for photon emission at frequency
$\omega$ 
\begin{multline}
\label{connectingshell}
w_{\uparrow g,gg}^{+}(\omega)=
\Gamma_{\rm ph}\frac{|{\tilde \Delta}_{h}||v_{e}|^2}{\pi(\varepsilon_{g}^{e}-E_{e}-\hbar{\tilde \omega})}\\\times\frac{\Theta(\varepsilon_{g}^{e}-E_{e}-\hbar{\tilde \omega}-|\Delta_{h}|)}{\sqrt{(\varepsilon_{g}^{e}-E_{e}-\hbar{\tilde \omega})^2-|\Delta_{h}|^{2}}}.
\end{multline}
Note that the spectrum of ``red" photons contains a continuous tail (see Fig.~3a) of main text) due to these processes. 
For the master equation, we need the total emission rate $W_{\uparrow g,gg}^{+}\equiv \int d\omega\, w_{\uparrow g,gg}^{+}(\omega)$, with the result 
\begin{equation}
\label{connectingtot}
W_{\uparrow g,gg}^{+}=
(\Gamma_{\rm ph}/\hbar)\,|v_{e}|^2\left|\frac{{\tilde \Delta}_{h}}{2\Delta_{h}}\right|.
\end{equation}
We remark that these cycle connecting rates also allow the population of triplet QD states, $|\uparrow\rangle_{e}|\uparrow\rangle_{h}$ and $|\downarrow\rangle_{e}|\downarrow\rangle_{h}$ and within our model are also responsible for the decay of triplet QD states that cannot proceed by direct recombination owing to selection rules (see Eq.~(15)).
Similar results as in Eqs.~(\ref{REP1}), (\ref{ROP1}), (\ref{connectingshell}) and (\ref{connectingtot}) hold for all processes that are included in the master equation [Eqs.~(17)-(32)].
In the cycle connecting processes, we only include terms to leading order in powers of $1/\Delta_{e,h}$. 
\begin{figure}[h]
\vspace{0.6cm}
\begin{center}
\includegraphics[width=0.8\columnwidth]{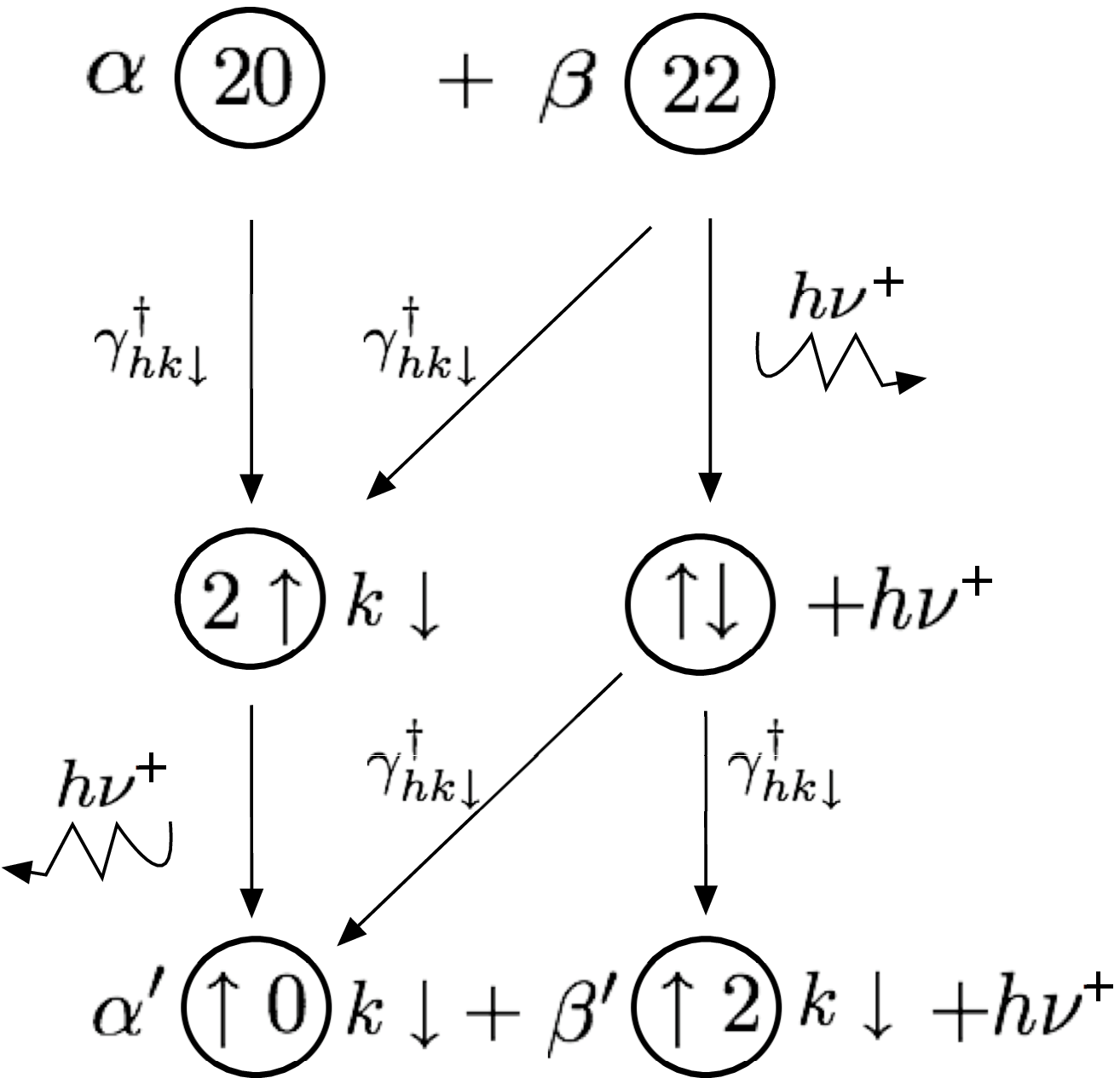}
\caption{Cycle connecting processes: Sketch of the cycle connecting process (proceeds in direction of arrows) that switches the QD state from the EP cycle with electron and hole singlets to $|\uparrow\rangle_{e}|g\rangle_{h}$ or $|\uparrow\rangle_{e}|e\rangle_{h}$ from the OP cycle (see also Fig.~\ref{scheme1}) and which creates a quasiparticle ($\gamma_{hk\downarrow}^{\dagger}$) in the SC lead (hole-side) with momentum ${\bf k}$ and spin $\downarrow$ thereby emitting a ``red" photon with frequency $\nu$ and circular polarization $+$. Only the initial state component with two electrons will take part in the process, whereas both components of the hole singlet  ($|0\rangle_{h}$ or $|2\rangle_{h}$) can participate in the process. The coherence factors $\alpha,\beta,\alpha',\beta'$ depend on specific QD states (initial and final states) involved. This diagram is therefore also relevant for the transition $|g\rangle_{e}|g\rangle_{h}\rightarrow |\uparrow\rangle_{e}|g\rangle_{h}$ discussed in the text. }
\end{center}
\label{connecting}
\end{figure}
\subsection{Spontaneous emission of ``red" photons}
We solve the master equation (Eqs.~(17)-(32)) in the stationary limit ${\dot \rho}_{a}=0$ and calculate the emission intensity according to
\begin{equation}
i(\omega)=\sum_{ab,p}w_{b,a}^{p}(\omega)\rho_{a},
\end{equation}
which leads to the plots of Fig.~3 in the main text, and to Fig.~1b) in this supplementary.

We now discuss the case of normal leads. When the leads are normal conducting, {\it single} particle transfer between the leads and QD is possible and the QD levels acquire the usual broadening ($\Gamma_{{\rm t};e}$ for the electron level and $\Gamma_{{\rm t};h}$ for the hole level). Since we are working in the parameter regime $\Gamma_{{\rm t};e},\Gamma_{{\rm t};h}\gg\Gamma_{\rm ph}$, the level broadening due to the electron hole recombination can be neglected here. This then leads to the following emission intensity (for $U=0$) at frequency $\omega$
\begin{multline}
\label{normal}
i_{\rm N}(\omega)=2\Gamma_{\rm ph} \int\limits_{-\infty}^{+\infty}d\varepsilon \int\limits_{-\infty}^{+\infty}d\varepsilon' f_{e}(\varepsilon)\nu_{e}(\varepsilon)[1-f_{h}(\varepsilon')]\\
\times \nu_{h}(\varepsilon')\delta(\varepsilon-\varepsilon'-\hbar\omega),
\end{multline}
where $f_{e,h}(\varepsilon)=(1+\exp[\beta(\varepsilon-\mu_{e,h})])^{-1}$ are the usual lead Fermi functions with $\beta=(k_{B}T)^{-1}$ and $\nu_{e,h}=(1/\pi)(\Gamma_{{\rm t};e,h}/2)/[(\varepsilon-E_{e,h}+\mu_{e,h})^2+(\Gamma_{{\rm t};e,h}/2)^2]$ are the DOS for the broadened QD levels. The factor 2 in Eq.~(\ref{normal}) accounts for the two polarizations of the emitted light. In Fig.~3a) of the main text we show (dashed curve) $i_{\rm N}(\omega)$ at zero temperature ($T=0$) and at resonance $E_{e}=E_{h}=0$ with the result (assuming $\Gamma_{{\rm t};e}=\Gamma_{{\rm t};h}$) $i_{\rm N}(\omega)=(\Gamma_{\rm ph}/|{\tilde \Delta}|)(2/\pi^2)F(\hbar{\tilde\omega}/|{\tilde \Delta}|)$ where $F(z)=\Theta(-z)\int_{0}^{-z}dx[(x^2+1)((x+z)^2+1)]^{-1}$. We note that the three discrete peaks in Fig.~3a) of the main text (SC case) approximately corresponds to the integrated emission intensity in the normal case and the leading contributions for large negative ${\tilde \omega}$ (continuous contribution in SC case) fall off in both cases like $1/{\tilde \omega}^2$.
\subsection{Spontaneous emission at the Josephson frequency $2eV_{\rm sd}/\hbar$}
Here, we describe the process of photon emission at the Josephson frequency $2eV_{\rm sd}/\hbar$. If only one photon per Cooper pair transfer from the $n$ side to the $p$-side is emitted it must have an energy $\sim 2eV_{\rm sd}$ which we call a ``blue" photon. Since the Cooper pair has charge $2e$ one electron-hole pair has to recombine without the emission of a photon which becomes possible in the presence of an externally applied dc electric field ${\bm E}_{0}$.
We now derive the form of the relevant Hamiltonian Eq.~(7) of the main text. 

Besides the Hamiltonian $H_{\rm int,1}$ that emits or absorbs photons, 
there is an additional part of the Hamiltonian related to a dc-electric field ${\bm E}_{0}$ 
\begin{equation}
\label{eq2}
H_{\rm int, 0}=-e\,{\bm r}\cdot{\bm E}_{0},
\end{equation}
where we assume that the field is homogeneous. Since the field is static it cannot provide photons.
In second quantization, $H_{\rm int, 0}$ reads
\begin{equation}
\label{eq3}
H_{\rm int, 0}=\sum_{\sigma\sigma'}V_{0}^{\sigma' \sigma}\,b_{\sigma'}^{\dagger}c_{\sigma}e^{-ieV_{\rm sd}t}+\rm{H.c.},
\end{equation}
with $V_{0}^{\sigma' \sigma}=\langle 0|b_{\sigma'}(-e {\bm r}{\bm E}_{0})c_{\sigma}^{\dagger}|0\rangle$.
Here, $c_{\sigma}$ and $b_{\sigma}$ denote annihilation operators for electrons with spin $\sigma$ for the level in the conduction band and valence band (their energies are again counted from respective chemical potentials $\mu_{e,h}$), respectively.

To calculate the matrix element $V_{0}^{\sigma'\sigma}$, it is crucial to know the orbital angular momentum of states that are connected by the operator Eq.~(\ref{eq2}). For QDs, usually the conduction band ground state level is an s-state ($l=0$) and the valence band state is a heavy hole p-state ($l=1$). Both states are two-fold degenerate with opposite total angular momentum $j_z$ in $z$-direction. For the conduction band these are $|1/2,+1/2\rangle$ and $|1/2,-1/2\rangle$, and for the valence band they are $|3/2,+3/2\rangle$ and $|3/2,-3/2\rangle$. Note that valence band states with either total angular momentum 3/2 but projection $\pm 1/2$ (light-holes) or the split-off band with total angular momentum 1/2 are lower energy states and are therefore occupied and far away from resonance with the leads.

The wave functions for the $s$-state in the conduction and for the heavy-hole (HH) $p$-state in the valence band are written in the envelope approximation as $\langle {\bm r}|c_{\sigma}^{\dagger}|0\rangle\simeq \phi_{c}(\bm r)u_{c\sigma}(\bm r)|\sigma\rangle$ and $\langle {\bm r}|b_{\sigma}^{\dagger}|0\rangle\simeq \phi_{v}(\bm r)u_{{\rm HH}\sigma}(\bm r)|\sigma\rangle$, respectively. Here $\phi_{c,v}$ are the envelope functions for the electron level and hole level, respectively, and $u_{c\sigma}(\bm r)$ and $u_{{\rm HH}\sigma}(\bm r)$ are the ${\bm k}=0$ ``Bloch"-parts of the wave functions which have the periodicity of the lattice. 
They reflect the symmetry of the band and have the form $u_{c\sigma}(\bm r)=R_{c}(r) Y_{0}^{0}(\theta,\phi)|\sigma\rangle$, $u_{{\rm HH}\uparrow}(\bm r)=R_{v}(r) Y_{1}^{+1}(\theta,\phi)|\uparrow\rangle$ and $u_{{\rm HH}\downarrow}(\bm r)=R_{v}(r) Y_{1}^{-1}(\theta,\phi)|\downarrow\rangle$. The spherical harmonics are 

\begin{equation}
Y_{0}^{0}(\theta,\phi)=1/\sqrt{4\pi},
\end{equation}

\begin{equation}
Y_{1}^{1}(\theta,\phi)=-\frac{1}{2}\sqrt{\frac{3}{2\pi}}e^{i\phi}\sin\theta,
\end{equation}
and
\begin{equation}
Y_{1}^{-1}(\theta,\phi)=\frac{1}{2}\sqrt{\frac{3}{2\pi}}e^{-i\phi}\sin\theta .
\end{equation}

To calculate $V_{0}^{\sigma' \sigma}$, we make use of the fact that the envelope part of the wave functions vary slowly on the 
scale of the lattice and write ${\bm r}={\bm r}_{i}+{\bm R}_{i}$ with ${\bm R}_{i}$ the Bravais lattice vector of the $i$-th unit cell \cite{Yoshibook}. Using the orthogonality of the periodic parts of the wave functions from the conduction band and valence band, we obtain
$V_{0}^{\sigma' \sigma}=V_{0}^{\sigma}\delta_{\sigma',\sigma}$ with
$V_{0}^{\sigma}={\bm d}_{\sigma}\cdot{\bm E}_{0}$. Explicitly, the interband dipole moment of the QD is ${\bm d}_{\sigma}=d(\sigma{\bm e}_{x}-i{\bm e}_{y})/\sqrt{2}$, where
\begin{equation}
\label{dipoleamplitude}
d=\frac{e}{\sqrt{3}}\sum\limits_{i}\phi_{\nu}^{*}({\bm R}_{i})\phi_{c}({\bm R}_{i})\int\limits_{0}^{R_{\rm W}}dr\,r^3R_{v}^{*}(r)R_{c}(r),
\end{equation}
with $R_{\rm W}$ the radius of the Wigner-Seitz cell (assumed to be sperical for definiteness). The amplitude of the dipole moment Eq.~(\ref{dipoleamplitude}) depends on the specific form and material of the QD. 
To transform to hole operators we replace $b_{\sigma}$ by $h_{-\sigma}^{\dagger}$ in Eq.~(\ref{eq3}) which leads to Eq.~(7) of the main text.
\begin{figure}[h]
\begin{center}
\includegraphics[width=0.7\columnwidth]{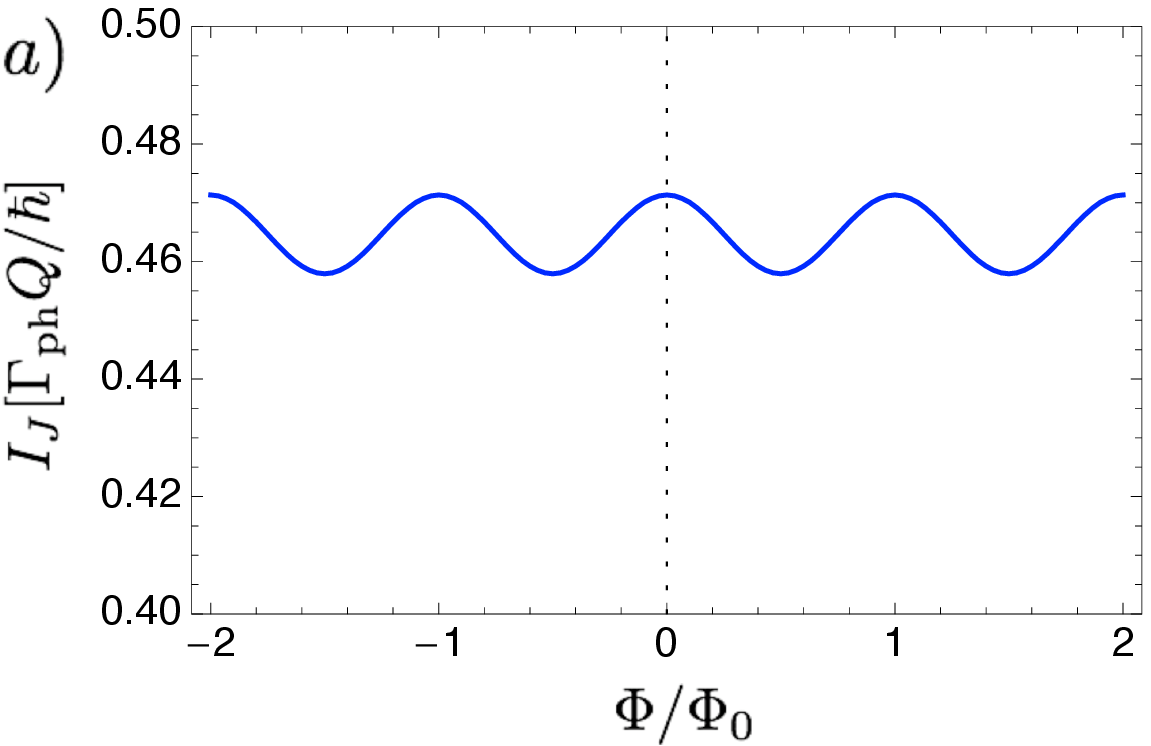}
\end{center}
\begin{center}
\includegraphics[width=0.62\columnwidth]{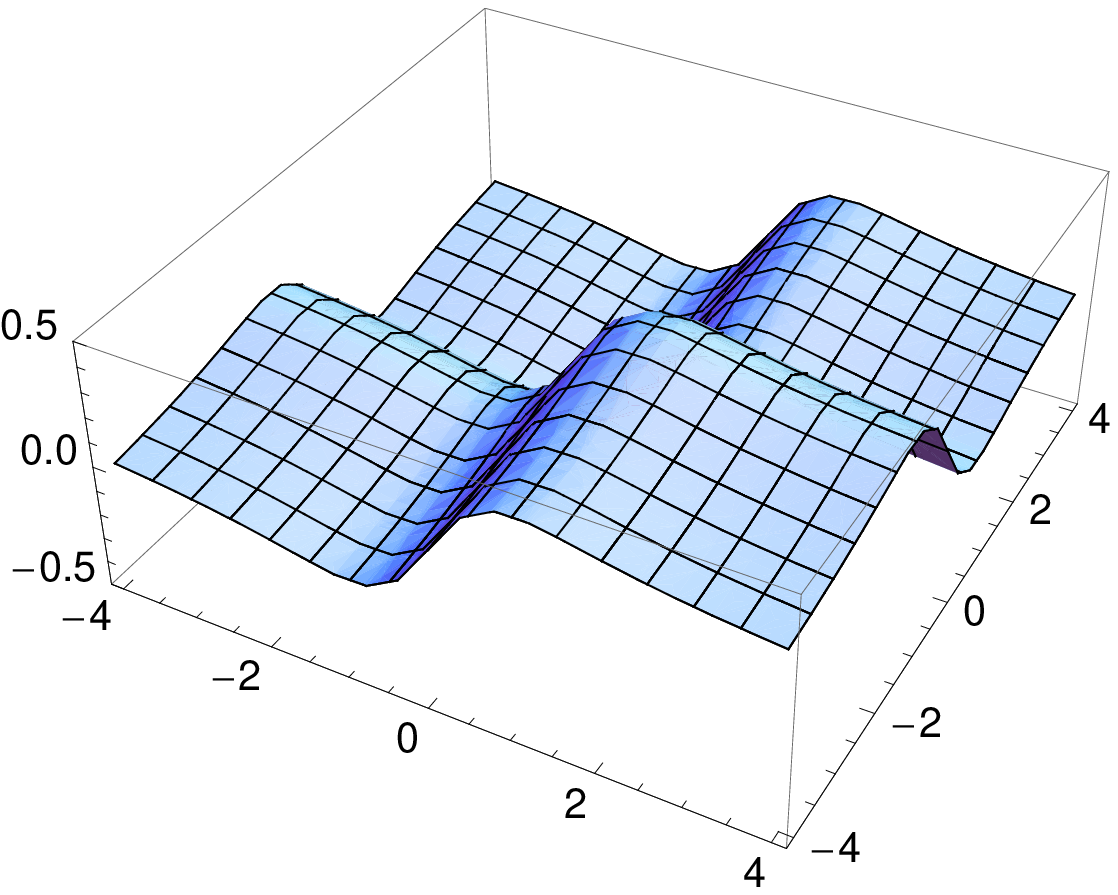}
\put(-112,7){$E_{e}/|{\tilde \Delta}|$}
\put(-13,34){$E_{h}/|{\tilde \Delta}|$}
\put(-165,120){$b)$}
\put(-165,34){\rotatebox{90}{$I_{J}^{\rm int}[\Gamma_{\rm ph}Q/\hbar]$}}
\end{center}
\begin{center}
\includegraphics[width=0.7\columnwidth]{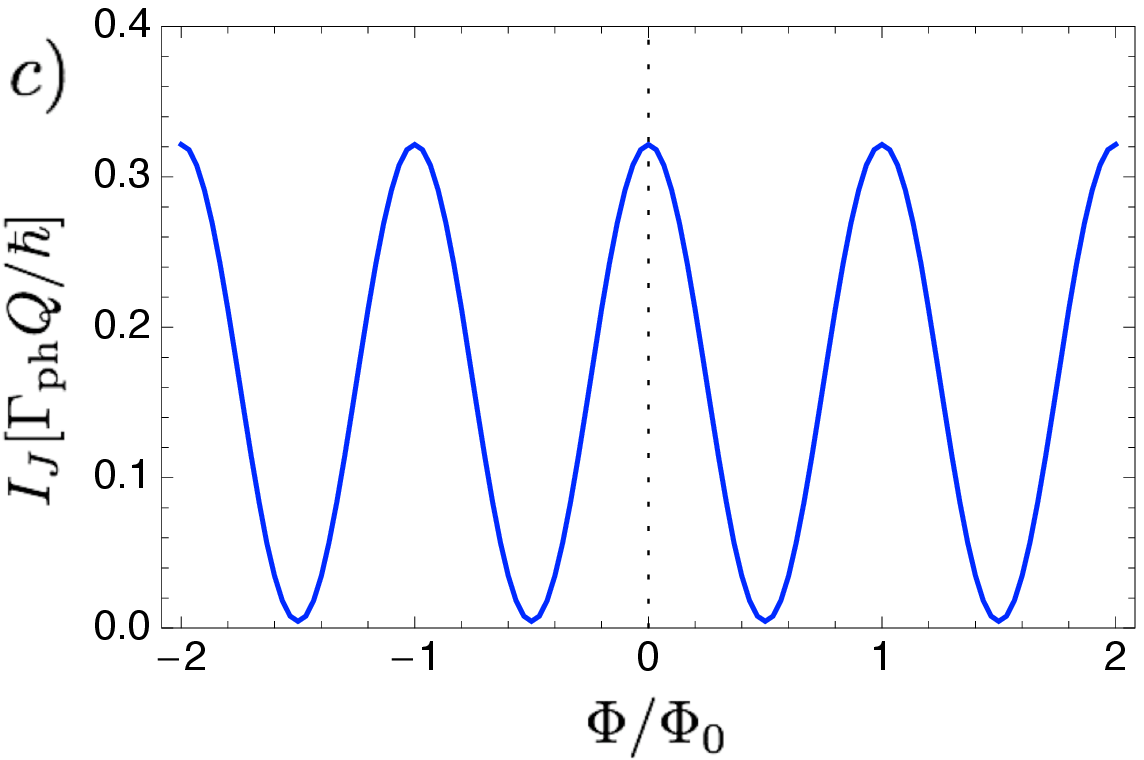}
\end{center}
\vspace{-0.4cm}
\caption{(Color online)
Emission intensity of proposed Josephson light-emitting diode in a SQUID configuration with two QDs (1,2) [see Fig.~4a) of main text] enclosing flux $\Phi$.
a) Generic (asymmetric) situation with $E_{e}=0.2$, $E_{h}=0.18$, $U=0.25$ in QD1; $E_{e}=0.1$, $E_{h}=-0.3$, $U=0.15$ in QD2. b) Dependence of interference part of $I_{J}$ ($I_{J}^{\rm int}$) at $\Phi=0$ as function of $E_{e}$ and $E_{h}$ in QD1 fixing $E_{e}=1$, $E_{h}=-1$ for QD2 ($U=0$ in both QDs). c) Regime where QDs are mainly in the state of ground state singlets $|g\rangle_{e}|g\rangle_{h}$ with $E_{e}=2.8$, $E_{h}=-3.1$, $U_{e}=0.23$ in QD1, $E_{e}=2.9$, $E_{h}=-3.2$, $U_{h}=0.6$ in QD2,  where almost destructive interference can be reached. 
For plots we choose $|{\tilde \Delta}_{1;e,h}|=|{\tilde \Delta}_{2;e,h}|\equiv |{\tilde \Delta}|$, equal optical path lengths from QDs 1,2 to D, energies in units of $|{\tilde \Delta}|$ and define $Q=2|V_{0}^{p}{\tilde \Delta}|^2/(eV_{\rm sd})^4$.
\label{SQUID}
}
\end{figure}
Having clarified the process of electron-hole recombination without photon emission we now can calculate the emission rates for emission of a single ``blue" photon, treating the total interaction Hamiltonian $H_{\rm int }\equiv H_{\rm int,0 }+H_{\rm int,1 }$ as a perturbation. 
We first discuss the emission of a ``blue" photon out of the EP cycle (see Fig. 2  upper cycle). Starting in the {\it singlet subspace} of electrons and holes, the emission process can be considered as a second-order ``emission'' process where the subspace of the doublet states $\{|\uparrow\rangle_{e}|\downarrow\rangle_{h},|\downarrow\rangle_{e}|\uparrow\rangle_{h}\}$ can only be occupied virtually, and the real transition therefore connects the QD singlet subspace with itself and produces a photon of $\hbar\omega\simeq 2eV_{\rm sd}$ since a charge $2e$ (in form of a Cooper pair) has been transferred by this process from the n-side to the p-side of the setup. Note that one of the electron-hole pairs present in the initial state is annihilated by $H_{\rm int,0}$ and the other pair by the spontaneous emission of a ``blue'' photon. The annihilation due to the dc-field can therefore also be interpreted as a {\it stimulated emission} of a zero frequency photon.

To second-order in time-dependent perturbation theory in $H_{\rm int }$, the emission rate of a ``blue" photon with polarization $p$ has the usual form \cite{Sakurai}
\begin{multline}
\label{bluerate}
W_{f,i}^{2eV_{\rm sd},p}=\frac{2\pi}{\hbar}\left|\sum\limits_{m}\frac{\langle f|H_{\rm int}|m\rangle\langle m|H_{\rm int}|i\rangle}{E_{i}-E_{m}+eV_{\rm sd}}\right|^{2}\\
\times\delta(E_{f}-E_{i}-2eV_{\rm sd}),
\end{multline}
where $|i\rangle$, $|m\rangle$, and $|f\rangle$ are the initial, intermediate and final states of the QD (energies counted from the respective chemical potentials) and elm-environment (photons) in the absence of the electric field. Note that the explicit time-dependence of $H_{\rm int}$ is accounted for by the $eV_{\rm sd}$-terms in the rate Eq.~(\ref{bluerate}).

We are interested in the emission of coherent photons, i.e. emission processes that leave the QD state {\it unchanged} after the photon is released. Such photons are emitted at the Josephson frequency $2eV_{\rm sd}/\hbar$.
Therefore the rate Eq.~(\ref{bluerate}) can be written as $W_{f,i}^{2eV_{\rm sd},p}=(\Gamma_{\rm ph}/\hbar)|{\cal A}^{p}_{a}|^{2}$, where $a$ denotes the initial (and final) state $|a\rangle$ of the QD.
We derive the following amplitudes:
\begin{equation}
{\cal A}_{gg}^{p}=\Lambda^{p}\frac{2(E_{e}+E_{h}-\varepsilon_{g}^{e}-\varepsilon_{g}^{h})}{(eV_{\rm sd})^2},
\end{equation}
\begin{equation}
{\cal A}_{ee}^{p}=\Lambda^{p}\frac{2(E_{e}+E_{h}-\varepsilon_{ex}^{e}-\varepsilon_{ex}^{h})}{(eV_{\rm sd})^2},
\end{equation}
\begin{equation}
{\cal A}_{ge}^{p}=-\Lambda^{p}\frac{2(E_{e}+E_{h}-\varepsilon_{g}^{e}-\varepsilon_{ex}^{h})}{(eV_{\rm sd})^2},
\end{equation}
\begin{equation}
{\cal A}_{eg}^{p}=-\Lambda^{p}\frac{2(E_{e}+E_{h}-\varepsilon_{ex}^{e}-\varepsilon_{g}^{h})}{(eV_{\rm sd})^2}.
\end{equation}
We note that emission processes out of the doublet states $|\sigma\rangle_{e}|-\sigma\rangle_{h}$ turn out to be of higher order ($\propto 1/(eV_{\rm sd})^3$) and will be neglected. We defined $\Lambda^{p}=V_{0}^{p}G|u_{e}v_{e}u_{h}v_{h}|\exp[i(\phi_{e}-\phi_{h})]$.

Within the OP cycle, similar amplitudes exist:
\begin{equation}
\label{OP1}
{\cal A}_{g\sigma}^{p}=-\Lambda^{p}\frac{(\varepsilon_{g}^{h}-\varepsilon_{ex}^{h})}{(eV_{\rm sd})^2},
\end{equation}
\label{OP2}
\begin{equation}
{\cal A}_{e\sigma}^{p}=\Lambda^{p}\frac{(\varepsilon_{g}^{h}-\varepsilon_{ex}^{h})}{(eV_{\rm sd})^2},
\end{equation}
and
\begin{equation}
\label{OP3}
{\cal A}_{\sigma g}^{p}=-\Lambda^{p}\frac{(\varepsilon_{g}^{e}-\varepsilon_{ex}^{e})}{(eV_{\rm sd})^2},
\end{equation}
\begin{equation}
\label{OP4}
{\cal A}_{\sigma e}^{p}=\Lambda^{p}\frac{(\varepsilon_{g}^{e}-\varepsilon_{ex}^{e})}{(eV_{\rm sd})^2}.
\end{equation}
As can be seen from Fig.~2, the ``blue" photon emission out of the OP cycle involves virtual QD states $|m\rangle$ that are a product of a doublet state and a singlet state.

To calculate the emission intensity at the Josephson frequency $I_{J}$, we use the same master equation [Eqs.~(17)-(32)] since the occupation probabilities of the QD states are determined by the faster ``red" photon emission (see discussion below), i.e. $I_{J}=\sum_{a,p}W_{a,a}^{2eV_{\rm sd},p}\rho_{a}$. To test the coherence of such photons we suggest an interference experiment of photons emitted from either of {\it two} QDs arranged in a superconducting quantum interference device (SQUID), see Fig.~4 a) of the main text. The emission intensity at the detector $D$ is $I_{J}=(\Gamma_{\rm ph}/\hbar)\sum_{aa',p}\rho_{a}\rho_{a'}|{\cal A}_{1,a}^{p}e^{i2\pi l_1/\lambda_{J}}+ {\cal A}_{2,a'}^{p}e^{i2\pi l_2/\lambda_{J}}|^2$, where $\lambda_{J}=hc/2eV_{\rm sd}$ is the wave length of coherent light at the Josephson frequency and $l_{1}$ and $l_{2}$ are the respective path lengths from the QDs to the detector. The interference contribution $I_{J}^{\rm int}$ is proportional to $\sum_{aa',p}\rho_{a}\rho_{a'}{\rm Re}[{\cal A}_{1,a}^{p}({\cal A}_{2,a'}^{p})^{*}]$ with ${\rm Re}[{\cal A}_{1,a}^{p}({\cal A}_{2,a'}^{p})^{*}]\propto \cos[2\pi((l_1-l_2)/\lambda_{J}+\Phi/\Phi_{0})]$, where we use that  $\phi_{1e}-\phi_{1h}-(\phi_{2e}-\phi_{2h})=2\pi\Phi/\Phi_{0}$, with $\Phi$ the flux through the SQUID and $\Phi_{0}=hc/2e$ the SC flux quantum. 
In Fig.~4b) of the main text we show the emission intensity in a regime where we observe a maximal interference contribution. Note that exactly at resonance $E_{e}=E_{h}=U=0$ in both QDs, the interference contribution vanishes due to different signs in amplitudes [see Eqs.~(47)-(54)]. The interference contribution can be on the same order as the total emission intensity $I_{J}$ for a quite general parameter set [see Fig.~\ref{SQUID}a)]  and has the order of magnitude 
\begin{equation}
I_{J}\sim 2 (\Gamma_{\rm ph}/\hbar)|{\bm d}\cdot {\bm E}_{0}|^2|{\tilde \Delta}|^2/(eV_{\rm sd})^4.
\end{equation}
Therefore, the ``blue" photon emission intensity is approximately by a factor $|{\bm d}\cdot {\bm E}_{0}|^2|{\tilde \Delta}|^2/(eV_{\rm sd})^4$ smaller than the ``red" photon emission.
We note that the bias voltage $eV_{\rm sd}$ over both QDs in the SQUID is necessarily the same. This leads to certain constraints regarding the similarities of the two QDs. Gate voltages, however, could be used to tune the QD levels in the conduction and valence band into the close proximity to the SC reservoirs. Fig.~\ref{SQUID}b) shows the sensitivity of the coherent contribution ($I_{J}^{\rm int}$) to the change of electron and hole energies of one of the QDs (leaving the parameters of the other QD fixed). The plot shows that the interference contribution $I_{J}^{\rm int}$ changes on the scale of $|{\tilde \Delta}|$
and therefore is not very sensitive to spectroscopic differences of the two QDs. In particular, the two QDs need not be identical within the linewidth $\simeq \Gamma_{\rm ph}\ll |{\tilde \Delta}|$. This linewidth (or broadening) of the $2eV_{\rm sd}$ emission is determined by the much faster ``red" photon emission which switches between different QD states. Fig.~\ref{SQUID}c) shows the regime where the
QDs are mostly in a single state ($|g\rangle_{e}|g\rangle_{h}$ for both QDs) and almost destructive interference can be reached.
Finally we note that the present phenomena are inherently of SC (Josephson) origin, since $I_{J}=0$ if $|u_{e}u_{h}v_{e}v_{h}|=0$ which is the case if ${\tilde \Delta}_{e}=0$ or  ${\tilde \Delta}_{h}=0$.


\begin{references}
\bibitem{Which_direction} 
This holds for photons propagating in $z$-direction. The relation is easy to
adjust for any propagation direction.
\bibitem{temperature}
The emission peaks are broadened by $\sim \hbar/\Gamma_{\rm ph}$, even for finite temperature $T$ as long as $k_{B}T\ll |\Delta_{e,h}|$.
\bibitem{Matrixelement}
For simplicity of the discussion, we assume $V_{0}^{\pm}$, $G$ and $|{\tilde \Delta}|$ to be the same in both QDs.
\end{references}
\begin{references}
\bibitem{ph_direction} 
Strictly speaking, this holds only
for photons propagating in $z$-direction. The relation is easy to
adjust for any propagation direction.
\end{references}
\end{document}